%% file: 0_Main.tex
\documentclass[lettersize,journal]{IEEEtran}
\usepackage{amsmath,amsfonts}
\usepackage{algorithmic}
\usepackage{algorithm}
\usepackage{multirow}
\usepackage{xcolor}
\usepackage{array}
\usepackage[caption=false,font=normalsize,labelfont=sf,textfont=sf]{subfig}
\usepackage{textcomp}
\usepackage{stfloats}
\usepackage{url}
\usepackage{verbatim}
\usepackage{graphicx}
\usepackage{cite}
\usepackage{hyperref}
\usepackage{tcolorbox}
\usepackage{enumitem}
\usepackage{soul}

\newcommand{\find}[1]{
\begin{tcolorbox}[leftrule=1mm,toprule=0mm,bottomrule=0mm,left=1pt,right=2pt,top=2pt,bottom=2pt
]
\em #1
\end{tcolorbox}
}

\begin{document}

\title{\textit{Do Chase Your Tail!} Missing Key Aspects Augmentation in Textual Vulnerability Descriptions of Long-tail Software through Feature Inference}

\author{Linyi Han, Shidong Pan, Zhenchang Xing, Jiamou Sun, Sofonias Yitagesu, Xiaowang Zhang, Zhiyong Feng
\thanks{Manuscript received XXX XXX, 20XX. (Corresponding author: Xiaowang Zhang)}
\thanks{Linyi Han, Sofonias Yitagesu, Xiaowang Zhang, and Zhiyong Feng are with the College of Intelligence and Computing, Tianjin University, Tianjin, China. e-mail: \{hanly2, xiaowangzhang, zyfeng\}@tju.edu.cn and sofoniasyitagesu@yahoo.com.}
\thanks{Shidong Pan, Zhenchang Xing, and Jiamou Sun are with the CSIRO's Data61, Canberra, Australia. e-mail: \{Shidong.Pan, Zhenchang.Xing, Frank.Sun\}@data61.csiro.au}
\thanks{Linyi Han is also the Center of National Railway Intelligent Transportation System Engineering and Technology.}
\thanks{Shidong Pan is also with the Australian National University.}
}

\maketitle
\input{1_Abstract}

\input{2_Introduction}

\input{3_Formative_study}

\input{4_Approach}

\input{6_Evaluation}

\input{7_Discussion}

\input{9_Related_work}

\input{8_Conclusion}

\input{11_Acknowledge}
\input{12_Refer.bbl}

\bibliographystyle{IEEEtran}

\end{document}

%% file: 1_Abstract.tex
\begin{abstract}

Augmenting missing key aspects in Textual Vulnerability Descriptions (TVDs) is crucial for effective vulnerability analysis. For instance, in TVDs, key aspects include \textit{Attack Vector}, \textit{Vulnerability Type}, among others. These key aspects help security engineers understand and address the vulnerability in a timely manner.
For software with a large user base (non-long-tail software), augmenting these missing key aspects has significantly advanced vulnerability analysis and software security research. 
However, software instances with a limited user base (long-tail software) often get overlooked due to inconsistency software names, TVD limited avaliability, and domain-specific jargon, which complicates vulnerability analysis and software repairs.
In this paper, we introduce a novel software feature inference framework designed to augment the missing key aspects of TVDs for long-tail software. 
Firstly, we tackle the issue of non-standard software names found in community-maintained vulnerability databases by cross-referencing government databases with Common Vulnerabilities and Exposures (CVEs).
Next, we employ Large Language Models (LLMs) to generate the missing key aspects. However, the limited availability of historical TVDs restricts the variety of examples. To overcome this limitation, we utilize the Common Weakness Enumeration (CWE) to classify all TVDs and select cluster centers as representative examples.
To ensure accuracy, we present Natural Language Inference (NLI) models specifically designed for long-tail software. These models identify and eliminate incorrect responses. Additionally, we use a wiki repository to provide explanations for proprietary terms.
Our evaluations demonstrate that our approach significantly improves the accuracy of augmenting missing key aspects of TVDs for log-tail software from 0.27 to 0.56 (+107\%). Interestingly, the accuracy of non-long-tail software also increases from 64\% to 71\%. As a result, our approach can be useful in various downstream tasks that require complete TVD information.

\end{abstract}

%% file: 2_Introduction.tex
\section{Introduction}

%
%
\begin{figure}[t!]
\centering

\includegraphics[width=1\linewidth]{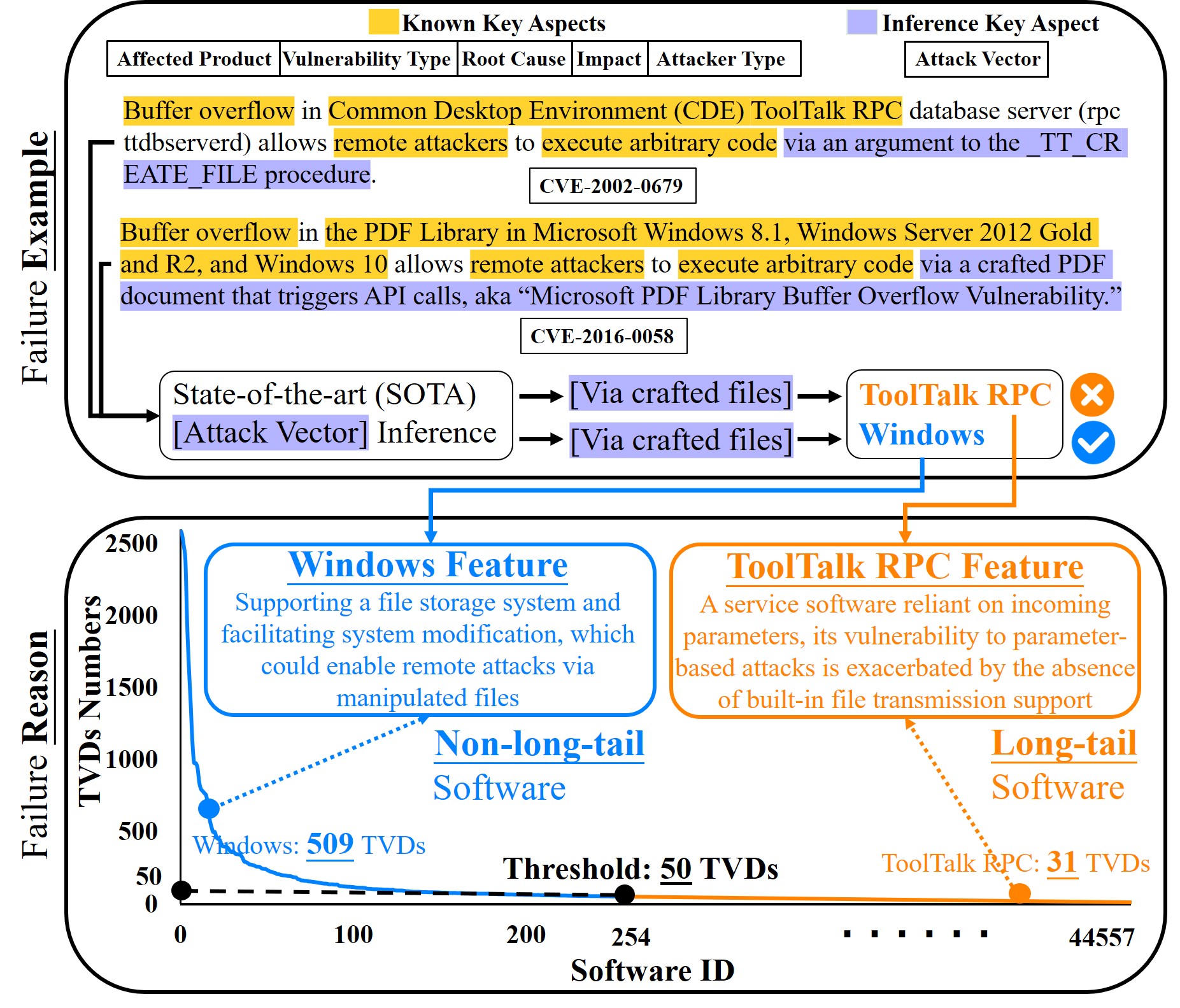}

\caption{In the coordinate axis, the x-axis represents the software instance represented by IDs, and the y-axis represents the number of TVDs each software has. For example, Windows, id=17, and has 509 TVDs. For CVE-2002-0679 (Microsoft Windows) and CVE-2016-0058 (ToolTalk RPC), we omitted the \textit{Attack Vector} information from TVDs. Microsoft Windows has 509 TVDs in the NVD, whilst  ToolTalk RPC only has 31 TVDs. 
To predict the missing key aspect (\textit{Attack Vector}) based on the rest of TVD's key aspects, none of the existing methods can successfully complete the task for ToolTalk RPC.
For Microsoft Windows, the vulnerability stems from file operations during runtime. It involves supporting a file storage system and facilitating system modification, which could enable attacks via manipulated files.
For ToolTalk RPC, a service software reliant on incoming parameters, its vulnerability to parameter-based attacks is exacerbated by the absence of built-in file transmission support.
Therefore, for SOTA predicting the same \textit{Attack Vector}, it is applicable to Windows but not to ToolTalk RPC.
}
\label{fig3}
\end{figure}

\IEEEPARstart 
{v}{ulnerabilities} are increasing in complexity and scale, posing great security risks to many software systems~\cite{DBLP:conf/iscc/WangZM23,DBLP:journals/stvr/LuckowKP20}. To address these risks, data sources such as the National Vulnerability Database (NVD)\cite{nvd} provide information on known vulnerabilities. They classify these vulnerabilities based on type and severity, providing common names, identifiers, links to patches, and additional details as short text descriptions.
In this paper, we focus on those informative short text descriptions of vulnerabilities, also known as TVDs, for software instances with a limited user base (referred to as ``long-tail software''). These software instances are often overlooked due to limited TVDs, variations in software features, and domain-specific jargon. By augmenting key aspects of TVDs, such as \textit{Vulnerability Type}, \textit{Attack Vector}, \textit{Attacker Type}, \textit{Impact}, and \textit{Root Cause}\cite{DBLP:conf/compsac/GuoXCLBZ21, JonathanEvans:2020}, we can improve vulnerability analysis~\cite{DBLP:conf/racs/SvacinaRWSCBSFT20,DBLP:journals/compsec/LiXZYC23}, software repairs~\cite{DBLP:journals/scn/ShenC20,DBLP:journals/corr/abs-2301-08653,DBLP:conf/uss/FengL0W0YZS19}, management\cite{DBLP:conf/eurosp/GeTPJ16, DBLP:conf/uss/BiswasFCRVNFP17}, mitigation\cite{DBLP:journals/tissec/PomonisPKPK18, DBLP:conf/ndss/WuHML20}, maintenance\cite{DBLP:conf/uss/LuPW19}, prevention\cite{DBLP:conf/icsm/HanLXLF17}, and other related tasks. TVDs are often incomplete on these key aspects, with a substantial missing rate ranging between 28\% and 42\%\cite{DBLP:journals/tosem/GuoCXLBS22}.
For example, CVE-2005-4676\footnote{\url{https://www.cve.org/CVERecord?id=CVE-2005-4676}} contains complete key aspects: ``Buffer overflow (\textit{Vulnerability Type}) in Andreas Huggel Exiv2 before 0.9 does not null terminate strings before calling the sscanf function (\textit{Root Cause}), which allows remote attackers (\textit{Attacker Type}) to cause a denial of service (application crash) (\textit{Impact}) via images with crafted IPTC metadata (\textit{Attack Vector})'', providing comprehensive information for software remediation efforts. In practice, however, certain key aspects might be missing. For example, if \textit{Attack Vector} is absent, the TVD becomes: ``Buffer overflow (\textit{Vulnerability Type}) in Andreas Huggel Exiv2 before 0.9 does not null terminate strings before calling the sscanf function (\textit{Root Cause}), which allows remote attackers (\textit{Attacker Type}) to cause a denial of service (application crash) (\textit{Impact})''. Without the \textit{Attack Vector}, software remediation teams may struggle to determine the attack path, making it difficult to devise an effective remediation strategy.
As a result, subsequent tasks, such as predicting software vulnerability level (CVSS)\cite{DBLP:conf/wcre/LiRXXS23}, identifying software vulnerability CWE category\cite{DBLP:journals/compsec/WangGRZ23}, and identifying vulnerability libraries\cite{DBLP:conf/iwpc/HaryonoK0SSA022}, commonly face challenges due to limited and incomplete training data, making it difficult to accurately extract features from TVDs.

In {Figure~\ref{fig3}}, we can see that non-long-tail software refers to widely used applications that have a large user base; a typical example is Microsoft Windows. On the other hand, long-tail software caters to niche industries or specific purposes and, therefore, has a smaller number of users. An example of long-tail software is ToolTalk RPC. 
To facilitate our analysis, we first examine the NVD data collected from 1999 to 2022. In line with previous studies~\cite{DBLP:conf/kbse/ZhouKXLHL23,DBLP:journals/ijcv/YangJSG22}, we utilize the Gini Coefficient\cite{ceriani2012origins} as a measure to quantify the long-tailedness. This measure, suggested by\cite{DBLP:conf/kbse/ZhouKXLHL23,DBLP:journals/ijcv/YangJSG22}, is insensitive to the number of data samples in the datasets. Our findings reveal that the distribution of software samples generally follows a long-tail pattern. This means that a small number of software instances have a significant number of TVD samples, which we refer to as ``non-long-tail'', while the remaining software exhibits extremely small numbers of TVD samples, referred to as the ``long-tail''.

Building upon previous research~\cite{DBLP:conf/kbse/ZhouKXLHL23}, we set a threshold of 50 TVDs (as shown in Table \ref{Threshold}). Software with fewer than 50 TVDs is labeled as ``long-tail'', while those with more than 50 TVDs are labeled as ``non-long-tail''. Based on this categorization, we have identified 44,303 long-tail software instances, which is 174 times greater than the number of non-long-tail software instances (254). Furthermore, the long-tail software types account for 97,753 TVDs, approximately 1.8 times more than those found in non-long-tail software (54,686). The analysis of long-tail software requires a deeper understanding of the domain, making vulnerability analysis, documentation, and security research more complex compared to non-long-tail software, which often has limited resources and documentation.

Previous research~\cite{DBLP:journals/tosem/GuoCXLBS22,DBLP:conf/wcre/LiRXXS23,DBLP:conf/iwpc/HaryonoK0SSA022,DBLP:journals/compsec/WangGRZ23} has shown that machine learning methods are effective in augmenting missing key aspects of TVDs for non-long-tail software (Section \ref{sec:Classification}). Additionally, pre-trained Large Language Models (LLMs)~\cite{openai} have demonstrated remarkable capabilities in various Software Engineering tasks~\cite{DBLP:conf/icse/DengXYZY024,DBLP:conf/icse/FengC24,DBLP:conf/sigsoft/00030L023, DBLP:conf/acl/ShapiraZG23}. However, when LLMs are directly applied to augment missing information in TVDs, their generative prediction performance for long-tail software is unsatisfactory (Section \ref{sec:formative_reasoning}). These methods utilize information from other TVDs of the same software to identify common patterns and make predictions. They have achieved good results on non-long-tail software with a large number of TVDs available for learning. However, as shown in {Figure~\ref{fig3}}, the distribution of TVDs among different software often exhibits an imbalance. Some software has significantly more samples (i.e., ``head''), while the majority have only a few samples (i.e., ``tail'').

[Inconsistency Software Name] The desired augmenting missing key aspects of TVDs for long-tail software faced significant challenges. 
Firstly, long-tail software is often developed within smaller, specialized communities. These communities frequently operate in isolation and adopt distinct naming conventions, resulting in significant variations in software nomenclature. For instance, “\textit{Struts REST}” (CVE-2017-9805) and “\textit{Struts2}” (CVE-2018-11776) reference the same software yet are labeled differently. The inconsistency in naming conventions makes it difficult to retrieve all TVDs associated with a specific software name.

[Limited TVD Availability] Secondly, long-tail software has a limited user base and low popularity. As a result, the volume of publicly available information related to these software instances is significantly restricted. This scarcity of data impedes the ability to identify relevant contextual knowledge or representative examples, which are essential for training neural networks, including large language models (LLMs). Consequently, these models struggle to learn meaningful representations for long-tail software, leading to reduced performance in previous studies~\cite{DBLP:journals/tosem/GuoCXLBS22,DBLP:conf/wcre/LiRXXS23,DBLP:conf/iwpc/HaryonoK0SSA022,DBLP:journals/compsec/WangGRZ23}.

[Domain-specific Jargon] Thirdly, long-tail software serves niche industries or specific purposes, which often leads to the use of domain-specific jargon, technical terms, and unique linguistic styles within vulnerability descriptions. Previous studies~\cite{DBLP:conf/compsac/GuoXCLBZ21, DBLP:conf/kbse/YitagesuXZ00H21,DBLP:journals/compsec/LiXZYC23} that relied on general-purpose language models struggled to capture and learn these specialized aspects accurately, thereby hindering their ability to accurately augment key aspects. Moreover, long-tail software exhibits various functionalities, platforms, and technologies, which poses a challenge for previous studies in identifying commonalities or shared patterns to generalize across different instances.

Lastly, the previous approach utilized the key information already available in TVDs for prediction. While this prediction method can be applied to a limited number of software (e.g., non-long-tail: 254), it faces challenges when there is a larger number of software instances (long tail: 44,303). In such cases, there can be two TVDs with the same key aspects but different missing key aspects due to different software features. The previous approach can only predict the same missing key aspects, limiting the effectiveness and reliability of their augmentation processes.

In this paper, we introduce a novel software feature inference framework designed to augment the missing key aspects of TVDs for long-tail software. To begin, we address the issue of non-standard software names that are often found in community-maintained vulnerability databases. These names can make it difficult to retrieve the TVDs based on the software name. We noticed that government-maintained databases, such as CNNVD~\cite{CNNVD} for China and CERT-FR~\cite{CERT} for France, use standardized software names that are different from those used by the community or international organizations. By cross-referencing these government databases with CVEs, we can cluster the TVDs under the same software name and gain richer information for subsequent analysis. 

Next, we utilize Large Language Models (LLMs) to generate missing key aspects in the TVDs using In-Context Learning. However, the scarcity of historical TVDs for long-tail software limits the diversity of examples. To overcome this, we employ the Common Weakness Enumeration (CWE)~\cite{cwe} to categorize all TVDs. By considering TVDs under the same CWE as candidate examples, we ensure a wider variety of representative examples. Inspired by clustering algorithms~\cite{DBLP:journals/pami/SelimI84}, we select cluster centers as examples, which further enhances their representativeness. This iterative process generates multiple
candidate answers and helps us leverage the valuable knowledge embedded within LLMs.

Finally, we rank the quality of the candidate answers and choose the best one as the final output for augmenting missing information. Large Language Models (LLMs)~\cite{openai} have limited knowledge of long-tail software, making them prone to generating incorrect responses. To address this issue, we integrate Natural Language Inference (NLI) models~\cite{DBLP:conf/iclr/GongLZ18}, specifically designed for long-tail software, to effectively identify and eliminate incorrect responses. The NLI model helps us establish the associative relationships between different key aspects in the TVDs. We filter the candidate answers based on the probability of association between software features and the candidate answers. However, the descriptions of product features often contain proprietary terms that have different meanings in the computer field. To overcome this challenge, we enhance the background information for product features by utilizing a wiki repository that explains these proprietary terms. Using a Learning Deep Structured Semantic Models (DSSM) model~\cite{DBLP:conf/cikm/HuangHGDAH13}, we embed both the software features and background information within the NLI model. The NLI results enable us to calculate the probability of association for
each candidate answers with all software features, and we select
the candidate answer with the highest probability as the final
augmented result.

We conducted extensive experiments to evaluate the effectiveness of our approach in different setups. The ablation experiments (RQ1 and RQ2) demonstrate that our software feature inference framework successfully addresses the issue of a long-tail distribution. Specifically, in the overall project comparison experiments (RQ3), the accuracy of augmenting missing key aspects for long-tail software increases from 0.27 to 0.56 (+107\%). Interestingly, the accuracy of non-long-tail software also improves from 0.64 to 0.71. Given its state-of-the-art performance, in the useful experiments (RQ4), our framework can be further applied to downstream tasks that rely on complete TVD information, such as predicting software vulnerability levels.

We have made the following contributions:
\begin{itemize} [leftmargin = *]
\item We conduct a formative study to explore the significance of the long-tail distribution of TVDs and identify challenges in current research that address these issues.
\item  We introduce a novel software feature inference framework designed to augment the missing key aspects of TVDs for long-tail software. Our framework enhances vulnerability information diversity and comprehensiveness by incorporating historical data and external knowledge sources.
\item We propose a method to detect hallucinations based on software features. Our method prioritizes individual software characteristics, allowing for exploring specific defect patterns unique to each software rather than adopting a universal approach to vulnerability pattern discovery.
\item We evaluate our approach using CVE (63,000), NVD (240), and NVD* (381) datasets, which demonstrate its effectiveness across various instances of long-tail software.

\end{itemize}

%% file: 3_Formative_study.tex
\hfill
\section{formative study}~\label{sec:formative}

We conduct a formative study to observe the challenges of long-tail software in vulnerability research.
The term \textit{long-tail} typically refers to the phenomenon where a significant portion of the data distribution comprises rare or infrequently occurring events or items. These long-tail instances often pose challenges for machine learning models because they have limited representation in the training data \cite{DBLP:conf/aaai/XuXLLJY23,DBLP:conf/kbse/ZhouKXLHL23,DBLP:conf/acl/DaiLZHL23}.
Despite LLMs' generalizability from large-scale pre-train, current methods rely on learning from existing information. However, due to the limited user base for long-tail software, historical TVDs are scarce, and the scarcity hinders effective parameter updates during model training and prevents learning long-tail software features.
In long-tail software, a distinct category is characterized by widespread adoption across numerous global IT sectors, such as \href{https://cn-ansibledoc.readthedocs.io/}{Ansible}.
However, due to limited number of TVDs (fewer than 50 TVDs), existing models still struggle to conduct effective vulnerability analysis. 

\subsection{Comparison of Existing Key Aspect Classification}
\label{sec:Classification}

Previous studies \cite{DBLP:journals/jss/SunYBWWZL23, DBLP:conf/icse/LyuLKWZLLL23} have focused on predicting key aspects using the Prediction of Missing Aspect (PMA) method \cite{DBLP:journals/tosem/GuoCXLBS22}. The PMA method is a neural network based classification model designed to predict missing key aspects based on existing key aspects in TVDs.
To evaluate the performance of the PMA method on different types of software, we compare its accuracy on both long-tail and non-long-tail software. First, we use the CVE dataset from Section~\ref{sec:dataset} for training, randomly selecting 4,000 entries as test data, and excluding these 4,000 entries from the training data. Following the data processing approach outlined in the original PMA paper, we randomly mask one key aspect in each TVD to serve as the ground truth. The remaining key aspects are used as input for the PMA model to predict the masked key aspect. The model utilizes a combination of CNN and LSTM, as this combination has been shown to outperform transformer models like BERT in the original PMA experiments. Finally, since PMA predefines labels for key aspects and maps them using word vector similarity, we apply the same method to map the ground truth to labels. We then compare the labels corresponding to the ground truth with those predicted by PMA, marking them as True if they match and False otherwise.

\begin{figure}[t!]
\centering
\includegraphics[width=1\linewidth]{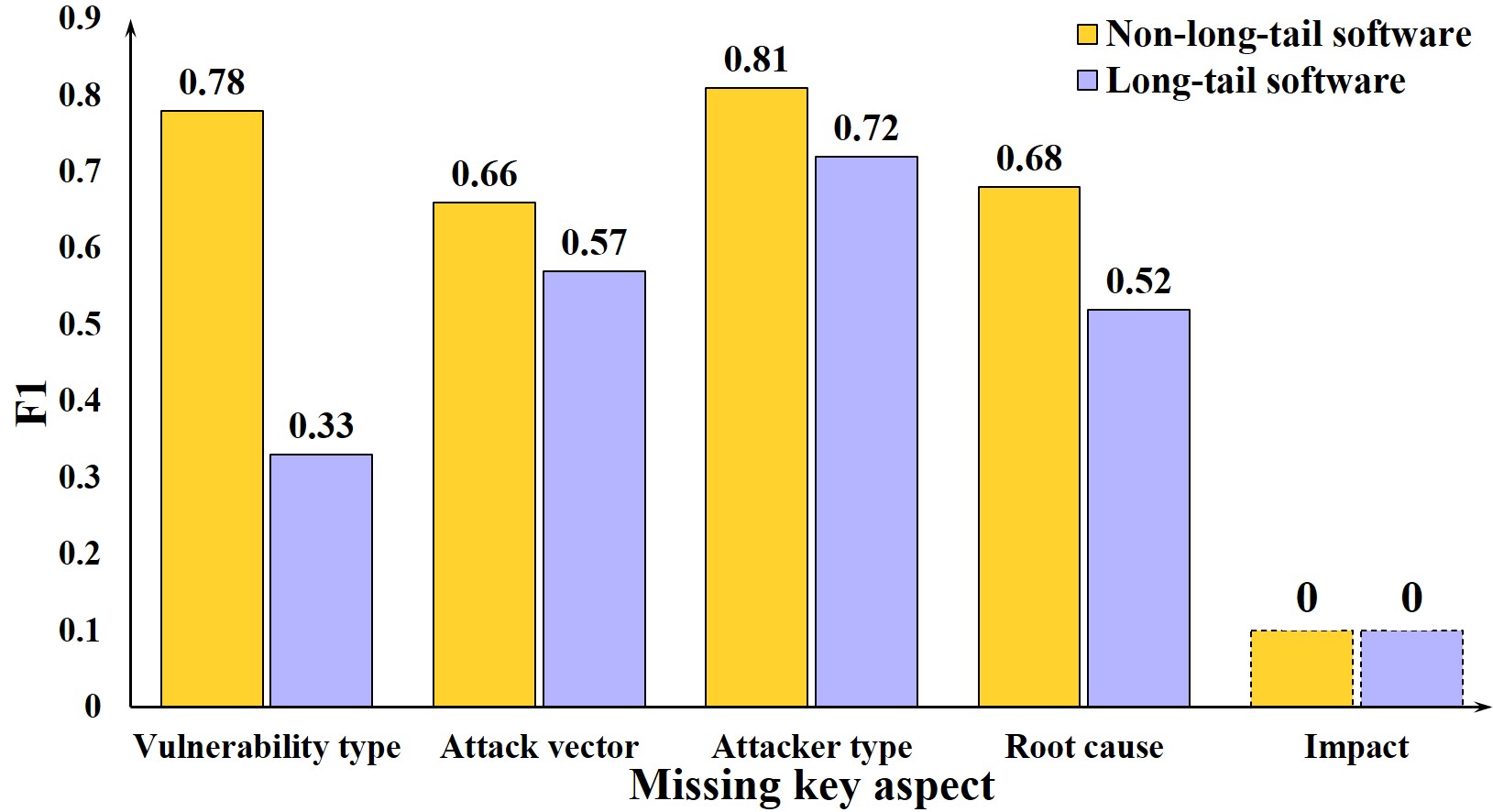}
\caption{Accuracy on classification task. The \textit{Impact} aspect has a value of 0, as PMA deems the missing rate of \textit{Impact} low and unnecessary to predict.}
\label{fig_clas}
\end{figure}

Figure \ref{fig_clas} presents the classification results of key aspects under different software types, and notably, the overall vulnerability type classification for long-tail software is distinctly lower than non-long-tail software. indicating distinct sensitivity levels to long-tail and non-long-tail software. The comparison highlights the difficulty in accurately classifying key aspects in TVDs for long-tail software.

\newcommand{\tabincell}[2]{\begin{tabular}{@{}#1@{}}#2\end{tabular}}

\begin{table}[]
  \centering
  \caption{F1 scores of different PMA structures. ``Original'' denotes the original PMA, ``BE.'' represents ``BERT'', ``LS.'' stands for LSTM, and ``LLa'' indicates ``LLaMA''.}
  \label{formative_tab}

  \begin{tabular}{c|cccccc}
  \hline
   \tabincell{c}{Soft.\\Type}&Structure&\tabincell{c}{Vuln.\\Type}& \tabincell{c}{Attack\\Vector}& \tabincell{c}{Attacker\\Type}& \tabincell{c}{Impact}& \tabincell{c}{Root\\Cause}\\
    \hline
    \multirow{4}{*}{\tabincell{c}{Long\\Tail}} 
    & \textbf{Original}& 0.33  & 0.57  & 0.72  & -- & 0.52 \\   
    & \textbf{BE.+LS.}& 0.36  & 0.55  & 0.69  & -- & {\bfseries0.56} \\ 
     & \textbf{BE.+BE.} & {\bfseries0.37}  & 0.55  & {\bfseries0.76}  & -- & 0.50\\  
     & \textbf{BE.+LLa.} & 0.31  & 0.52  & 0.71  & -- & 0.55 \\   
     & \textbf{LLa.+LLa.} &  {\bfseries0.37} & {\bfseries0.59}  & 0.68  & -- &  0.55\\      
    \hline
    \multirow{4}{*}{\tabincell{c}{Non\\Long\\Tail}} 
        & \textbf{Original}&  0.78 & 0.66  &  0.81 & -- & 0.68 \\   
    & \textbf{BE.+LS.}& 0.77  &  0.62 &  0.83 & -- & 0.61 \\   
     & \textbf{BE.+BE.} & 0.75  & 0.63  & 0.87  & -- & 0.63 \\   
     & \textbf{BE.+LLa.} & {\bfseries0.81}  & 0.61  & {\bfseries0.88}  & -- & 0.59 \\   
     & \textbf{LLa.+LLa.} & 0.78  & {\bfseries0.69}  &  0.85 & -- & {\bfseries0.71} \\  
    \hline
  \end{tabular}
\end{table}

To further investigate the potential impact of the capability of the backbone neural network in PMA toward the long-tail issue, we replace the Convolutional Neural Network (CNN) and Long Short-Term Memory (LSTM) in PMA with more advanced models and evaluated its performance. 
Table \ref{formative_tab} shows that although there is an improvement in PMA performance with more powerful and sophisticated neural network architectures (e.g., BERT and LLaMA), the increase is not significant and the long-tail issue still persists.

\subsection{Comparison of Key Aspect Generative Prediction}
\label{sec:formative_reasoning}
Previous research
primarily formulated the missing key aspect augmentation as supervised classification task. 
Thus, the performance is inevitably affected by the quantity and quality of training data, and long-tail software only has few valid training samples. 
With the recent advancements of generative models, such as pre-trained LLMs, generative prediction can offer more detailed information and achieve more desired performance compared to classification-based prediction.
Therefore, in this paper, we are the first to formulate the missing key aspect augmentation as a predictive generation task.
Pre-trained LLMs demonstrate outstanding capability on various Software Engineering tasks~\cite{DBLP:conf/icse/DengXYZY024,DBLP:conf/icse/FengC24,DBLP:conf/sigsoft/00030L023, DBLP:conf/acl/ShapiraZG23}; thus, we explore the feasibility of employing LLMs on predictive generation of missing key aspects in TVDs.

\begin{figure}[t!]
\centerline{\includegraphics[width=1\linewidth]{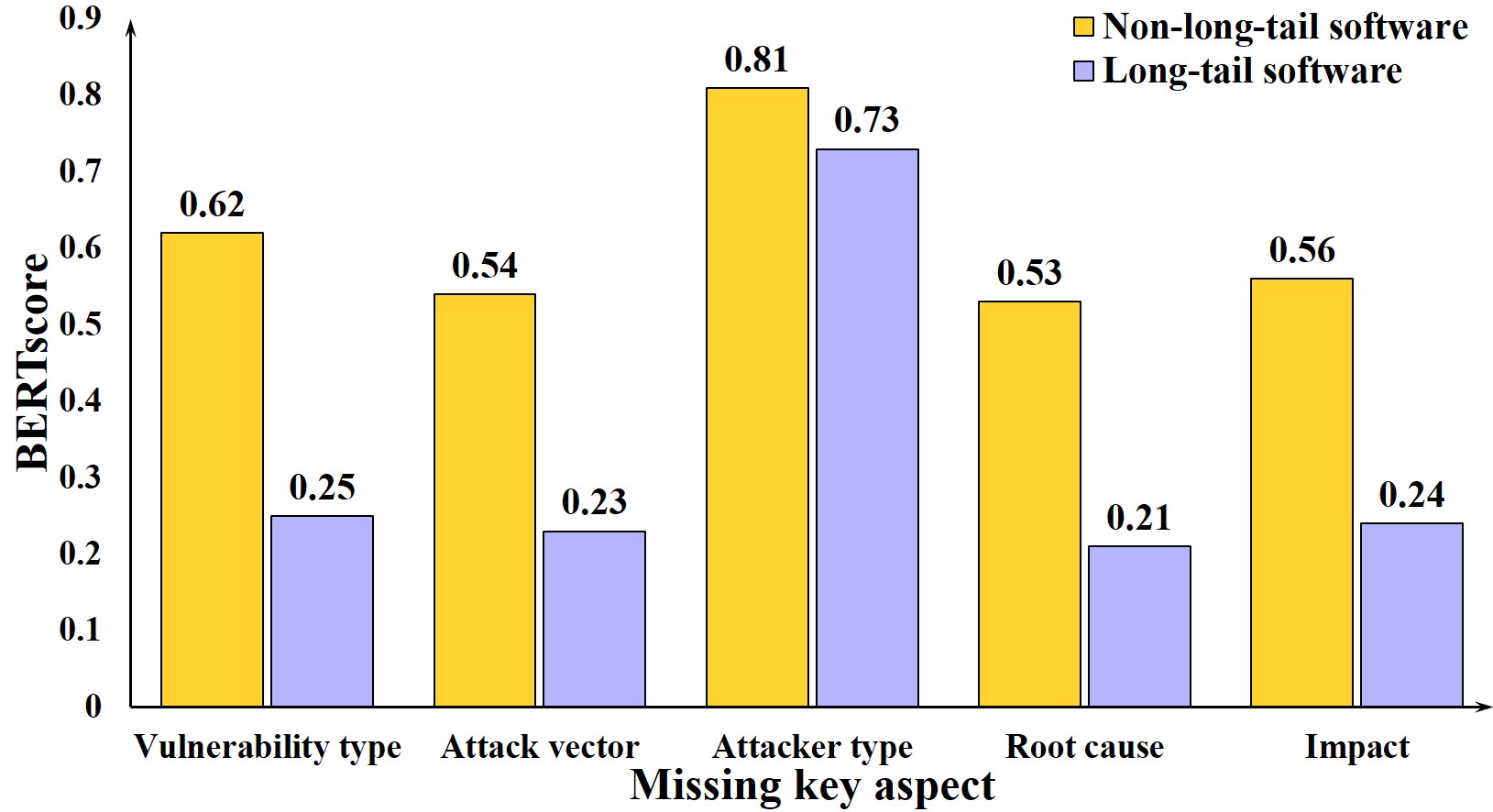}}
\caption{Performance of the generative prediction task on missing key aspect augmentation.}
\label{fig_inf}
\end{figure}

By plainly applying LLMs on augmentation, the generative prediction performance of long-tail software is still unsatisfying.
In the same experimental settings as described in Section~\ref{sec:setup}, Figure \ref{fig_inf} shows that the generative prediction performance of long-tail software is overall subpar compared to non-long-tail software.
Moreover, the disparity in BERTscore between different types of software is notably more pronounced in the generative prediction than classification-based prediction.
These observations underscore significant challenges posed by generative prediction for long-tail software.

It is broadly discussed that the prompt engineering will greatly affect the task performance of LLMs~\cite{DBLP:journals/corr/abs-2303-18223,DBLP:conf/nips/Wei0SBIXCLZ22,DBLP:conf/nips/KojimaGRMI22}.
Also, integrating external knowledge into the LLM-based frameworks can also boost the performance depending on the task~\cite{DBLP:conf/icse/DengXYZY024,DBLP:conf/icse/FengC24,DBLP:conf/acl/ShapiraZG23}.
In particular, by observing the fail cases (incorrect key aspect and hallucinatory explanation) of generative prediction of long-tail software, we find that the software features are prone to being overlooked by LLMs.
Therefore, software features is essential for improving the reasoning process, especially for long-tail software. We will delve into the definition and analysis of software features, highlighting their importance in enhancing the accuracy and coverage of LLM predictions.

%
%

\subsection{Software Feature}
\subsubsection{Definition of Software Feature}
According to classical software engineering literature, Sommerville's Software Engineering~\cite{sommerville2010software}, software features are specific functionalities or characteristics that meet user or system requirements. These features are identified through Natural Language Inference (NLI) models from large-scale textual data, such as Common Vulnerabilities and Exposures (CVE) reports. Our study uses NLI models to learn software features by analyzing recurring patterns and detailed descriptions in vulnerability description documents (TVDs). For example, the model identifies features like ``file storage systems'' and ``system modifications'' from terms such as ``file upload'' and ``permission management'' in Windows TVDs.

We also use external knowledge base to clarify ambiguous terms in the TVD dataset. For instance, ``file management'' in the TVD could include actions like file creation or permission management. External knowledge base's (e.g., Wikipedia~\cite{wiki}) detailed explanations help specify features such as ``NTFS file system'', ``file permissions'', and ``disk quotas'' enabling our NLI model to extract precise software features.

\subsubsection{Analysis of Software Feature}
In our study, we performed software feature analysis using a randomly selected sample of 400 vulnerability reports from the CVE dataset. These samples were processed using Natural Language Inference (NLI) models to identify recurring terms and patterns representing software features. The features were then categorized based on their frequency within the sampled reports. This random sampling approach ensures that the extracted features are representative of the broader dataset while allowing us to focus on the most relevant and distinguishing software characteristics.

The following a) and b) summarizes the most and least frequent software features identified from the 400 randomly sampled CVE reports:
\paragraph{Most Frequent Software Features}
Three software features appear most frequently in the CVE reports:
1. \textit{File Management}: Handling file operations such as NTFS, file permissions, and file creation/deletion.  
2. \textit{Network Access}: Managing network protocols and data transfer processes.  
3. \textit{User Authentication}: Facilitating user login credentials management and session handling.
\paragraph{Least Frequent Software Features}
Conversely, the following features are identified as least frequent:
1. \textit{Cryptography}: Performing encryption and decryption tasks. 
2. \textit{Multi-Factor Authentication}: Supporting additional authentication mechanisms beyond basic login credentials.  
3. \textit{Virtualization}: Enabling operations related to virtual machines and virtual environments.

These features, extracted from a random sample of the CVE dataset, reflect both common and specialized software functionalities. The most frequent features involve core system operations like file management and network access, widely found across various software applications. Conversely, the least frequent features, such as cryptography and multi-factor authentication, are associated with more specialized and security-focused tasks. This analysis allows our inference model to generalize effectively across different software instances, supporting accurate reasoning even for less common software vulnerabilities.

\subsubsection{Impact of Software Feature on Long-Tail Software}
This broad yet specific definition of ``software features'' supports scalable and accurate reasoning for long-tail software vulnerabilities. It allows us to address diverse software requirements by identifying both general operations (e.g., file management) and advanced operations (e.g., encryption). This hierarchical feature extraction method provides high coverage for non-long-tail software (like Windows) and accurately identifies unique traits in long-tail software (e.g., niche tools).

In long-tail scenarios, where training data is sparse, this approach is crucial. By defining broad yet functional features, our NLI model identifies generalizable functionalities in sparse data, improving performance on long-tail software. External knowledge base expands background knowledge, helping to address data scarcity by providing context for niche tools and clarifying ambiguous terms. This process enhances the model's generalization capability in long-tail scenarios.

Ultimately, this feature definition strategy improves our model's reasoning for long-tail software and reduces reliance on domain-specific knowledge, making our framework adaptable and accurate across various software types. Thus, our framework maintains consistent inference accuracy even with long-tail software.

%% file: 4_Approach.tex
\section{Framework}

This section outlines our software feature inference framework to augment missing key aspect in TVDs as generative prediction.
The framework comprises three essential modules to effectively address the challenges presented by the long-tail nature of software TVDs. 

\begin{figure*}
	\centering

	\includegraphics[scale=0.5]{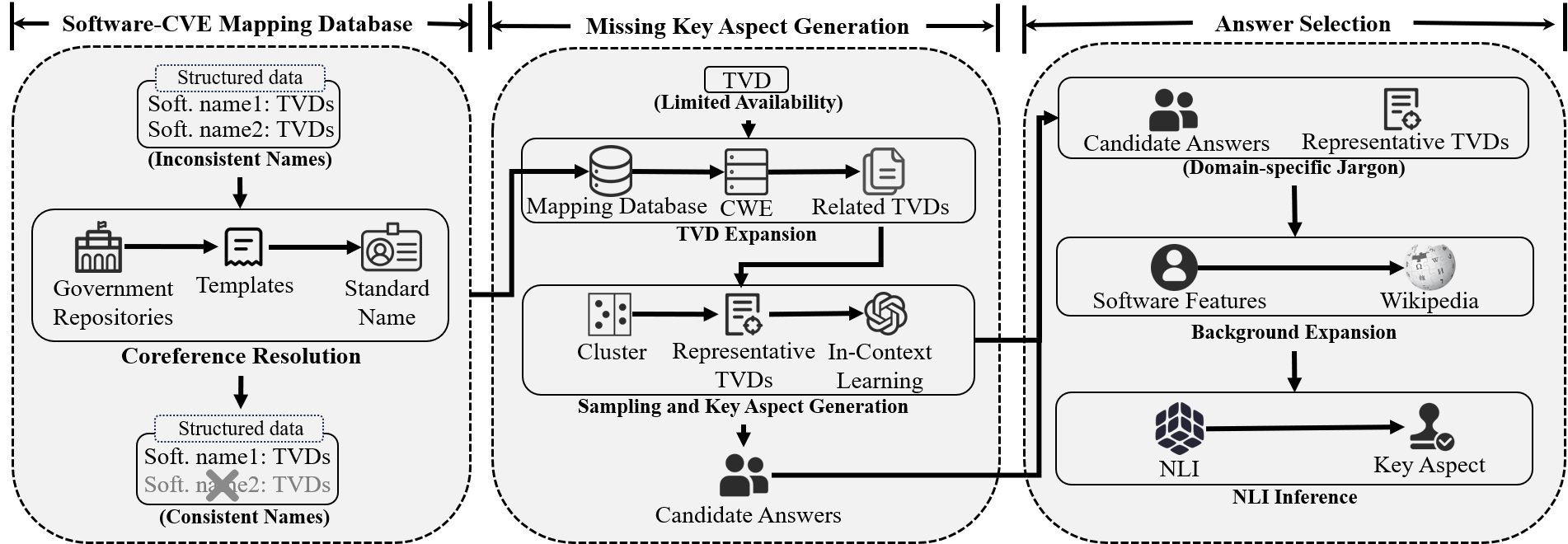}
	\caption{Approach overview: First, to construct software-CVE mapping database, we leverage government databases to resolve software naming references. Second, we utilize the CWE to enhance the diversity of historical TVDs associated with long-tail software. Lastly, in the face of the heightened risk of hallucination in LLMs due to the limited long-tail software knowledge stored, we employ software feature inference to mitigate this issue.}
	\label{fig2:Architecture}	
\vspace{-5mm}
\end{figure*}

\subsection{Software-CVE Mapping Database}

\label{Database}

Inconsistent software naming is commonplace in TVDs. Irregular software names can lead to different names with distinct TVDs referring to the same software, as observed in cases such as ``\textit{Struts REST}'' (CVE-2017-9805) and ``\textit{Struts2}'' (CVE-2018-11776), both referring to the same software. This phenomenon is particularly prevalent in long-tail software instances, where naming conventions vary widely.
Current research \cite{DBLP:journals/corr/abs-2202-01110, DBLP:conf/acm/AsaiMZC23, DBLP:journals/corr/abs-2306-06427} focused on constructing software name dictionaries and leveraging knowledge stored in LLMs for question answering. These methods disambiguate software names based on records of all the names a software has been called. However, long-tail software presents unique challenges in name disambiguation, as it is difficult for individuals or even LLMs to maintain familiarity with the numerous aliases associated with these software instances, rendering it impractical to list all the aliases of software exhaustively.

\begin{figure}[htbp]
    \centering
    \includegraphics[width=0.45\textwidth]{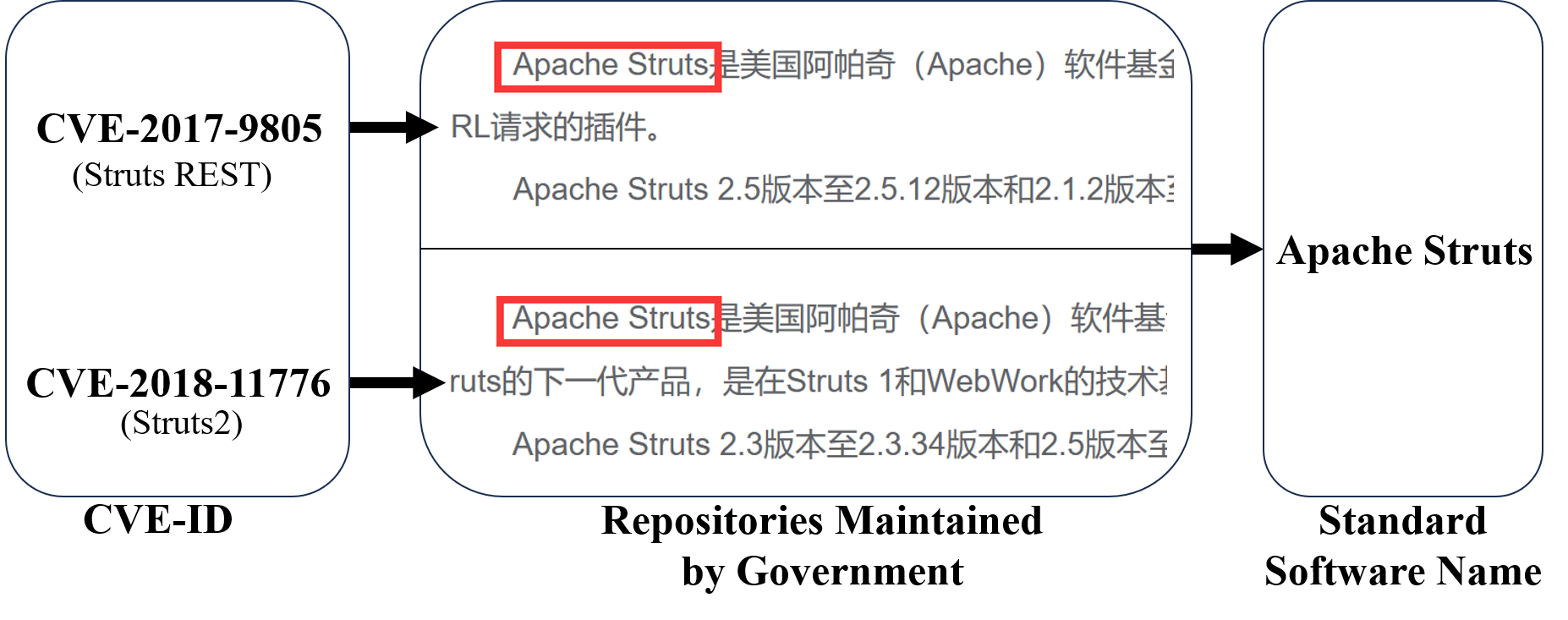}
    \caption{Software naming coreference resolution}
    \label{coreference}
\vspace{-5mm}
\end{figure}
Therefore, we first build the Software-CVE Mapping Database which contains all TVDs in CVE associated with their corresponding software, enabling efficient retrieval of all TVDs to a standard software name.
We discover that repositories maintained by government entities share common characteristics, such as the utilization of standardized software names and their linkage to TVDs in CVE through CVE-ID as the primary key in their databases. 
Thus, we opt to select three prominent government-maintained TVD repositories, CNNVD (China), JVN ipedia (Japan), and CERT-FR (France). These repositories rank among the top three with the highest number of TVDs, thereby offering a rich source of vulnerability information essential for achieving comprehensive coverage of software vulnerabilities.
Specifically, we first extract the standardized software names of TVDs from these repositories, and then link the software names with TVDs as the Mapping Database by the primary key CVE-ID. 
Notably, in those regional government-maintained repositories, TVDs contain both English and non-English content, but the software names are commonly written in English.
Also, we empirically find that TVDs typically starts with the corresponding software names at the beginning. 
Consequently, our extraction process specifically targets the English content in the first sentence of TVDs.
We delineate three templates for extracting English content from the repositories, tailoring for CNNVD, JVN ipedia, and CERT-FR, respectively.

\begin{itemize}[leftmargin=*]
\vspace{5pt}
    \item \text{template1:} \textit{[\textbackslash u4e00-\textbackslash u9fa5]([a-zA-Z]+[\textbackslash s-]?[a-zA-Z])} 
    \item \text{template2:}\textit{([a-zA-Z]+[\textbackslash s-]?[a-zA-Z])(?:\textbackslash d+(?:\textbackslash .\textbackslash d+)?[.])?}
    \item \text{template3:}\textit{[\textbackslash u4e00-\textbackslash u9fa5]([a-zA-Z]+)(?:[\textbackslash s-].*?)}
\vspace{5pt}
\end{itemize}

For instance, in Figure \ref{coreference}, the CVE-2018-11776 and CVE-2017-9805 are associated with ``\textit{Struts2}'' and ``\textit{Struts REST}'', respectively. 
Based on the Software-CVE Mapping Database we craft, those two names should be standardized as ``\textit{Apache Struts}'' according to the CNNVD.

We present a novel approach leveraging government-maintained TVD repositories to standardize software names. By designing repository-specific extraction templates, we achieve accurate name disambiguation. This mapping database not only addresses the naming inconsistency problem but also provides a unified, standardized dataset.

\subsection{Missing Key Aspect Generation}

\label{sec:Missing Key Aspect Reasoning}

In section~\ref{sec:formative}, we compare and discuss the classification and generative prediction to obtain missing key aspects. 
Directly applying LLMs on generative prediction cannot mitigate the under-performance for long-tail software.
In-context learning is a common strategy to increase the performance of LLMs on various code-related tasks~\cite{liao2023context}, which is providing several in-context few-shot learning examples as a part of the prompt.
Studies have shown that more relevant and more representative examples can significantly enhance the LLMs capability~\cite{DBLP:journals/corr/abs-2002-08909,DBLP:conf/iclr/ZhouSHWS0SCBLC23}.
However, previous studies~\cite{DBLP:journals/ieeesp/Massacci24, DBLP:journals/jss/WangLLJ24} did not differentiate between various software entities.
Different software entities exhibit significant differences in features. As shown in Figure~\ref{fig3}, Windows and ToolTalk RPC have distinct reasoning and prediction logic for key aspects. Therefore,
irrelevant software TVDs contribute noise to the prediction of target software TVDs.
The scarcity of historical TVDs for long-tail software leads to low diversity in examples, leaving us few selections for the in-context learning.

To obtain more relevant and representative TVD examples for the in-context learning, we employ the CWE to expand the historical TVDs of long-tail software, ensuring relevance of TVD examples. Furthermore, we adopt a method~\cite{DBLP:journals/eswa/LimLLT24} of selecting cluster centroids to ensure the diversity of TVDs, enhancing the representativeness of examples.
We formulate our method as depicted in Algorithm \ref{alg:alg3}, and the specific description of each step is elaborate below.

\subsubsection{Example Selection}
\label{Knowledge Retrieval}

To enhance the diversity of TVDs for long-tail software, we utilize the CWE. As depicted in lines 5-7 of Algorithm \ref{alg:alg3}.
We extract the CWE-ID corresponding to each TVD, which we obtain from the NVD database. Then, we extract all TVDs associated with the CWE-ID and merge them with the TVDs extracted from the Software-CVE Mapping Database. 
For instance, in the case of CVE-2002-0679 of Second module on Figure \ref{fig2:Architecture}, the corresponding CWE-ID is 119. While the software name of CVE-2002-0679 corresponds to only 6 TVDs, CWE-119 contains 207 TVDs. By combining these two sets of TVDs, a total of 213 TVDs are utilized.

\begin{algorithm}[t]
	\caption{Missing Key Aspect Generation}\label{alg:alg3}
	\begin{algorithmic}
    \STATE 1:\quad Let $data \text{ be the Software-CVE Mapping Database}$;
    \STATE 2:\quad Let $all\_names \text{ be the } [\text{``Vuln. Type''},\text{``Impact''},$
    \STATE $ \text{``Attack Vector''},\text{``Attacker Type''}, \text{``Root Cause''}] $;
    \STATE\quad $\cdots \cdots \cdots \cdots \cdots \cdots \cdots \cdots \cdots \cdots \cdots \cdots \cdots \cdots \cdots \cdots$
    \STATE 3:\quad $names \gets ExtractAspectName(\text{TVD})$;
    \STATE 4:\quad $m \gets all\_aspect\_names - names$;
    \STATE 5:\quad $X1 \gets GetKnowledge(v.\text{softwarename}, data);$
    \STATE 6:\quad $X2 \gets CWERelatedTVD(\text{TVD})$;
    \STATE 7:\quad $X \gets X1 + X2$;
    \STATE 8:\quad $E1 \gets WordVec([x_1, x_2, ..., x_n])$;
    \STATE 9:\quad $clu \gets Kmeans(E1, cluster\_number = 30)$;
    \STATE 10:\quad $k \gets ExtractKeyAspect(\text{TVD})$;
    \STATE \quad $\cdots \cdots \cdots \cdots \cdots \cdots \cdots \cdots \cdots \cdots \cdots \cdots \cdots \cdots \cdots \cdots$
    
    \STATE 11:\quad Let $c \text{ be the centers of } clu$;
    \STATE 12:\quad $direct\_generation \gets GenerationStruct(c, m, k)$;
    \STATE 13:\quad $fill\_in\_the\_blank \gets fillStruct(c, m, k)$;
    \STATE 14:\quad $selection \gets sellectStruct(c, m, k, $
    \STATE \hspace{0.5em}$LLM(direct\_generation), LLM(fill\_in\_the\_blank))$;
    \RETURN $LLM(selection)$
	\end{algorithmic}
 \label{alg:alg3}
\end{algorithm}

The introduction of CWE results in a large number of samples in prompt, as illustrated by Figure \ref{fig2:Architecture}, where the TVDs numbers of ToolTalk RPC and CWE-119 is 213. Therefore, it becomes essential to identify representative examples. Representative knowledge holds more significance \cite{DBLP:conf/acl-deelio/LiuSZDCC22, DBLP:conf/naacl/RubinHB22, DBLP:conf/acl/DoostmohammadiN23, DBLP:journals/corr/abs-2301-00303} than sheer volume, particularly highlighted in Section \ref{sec:RQ2}. Excessive knowledge can adversely impact prediction accuracy \cite{liu2023lost}. Table \ref{optimal_number} indicates 30 as the optimal CVEs for prediction.

If the number of retrieved TVDs is fewer than 30, we select all of them. However, if it exceeds 30, we address the overlength issue by employing clustering, as outlined in lines 8-9 of Algorithm \ref{alg:alg3}. Let $X={x_1,x_2,...,x_n}$ denote the set of retrieved TVDs, each associated with key aspects $C_{i}={c_{i1},c_{i2},...c_{im}}$. We utilize Word2Vec \cite{word2vec} to obtain sentence vectors $E={e_1,e_2,...,e_n}$. This involves training all CVE data using Word2Vec and generating vectors for each $x_i$. The Word2Vec sentence vector equals the sum of each word vector in the TVD. Next, we apply Kmeans \cite{DBLP:journals/tit/CoverH67} with 30 clusters, from which we extract TVDs of cluster centers.

\subsubsection{Candidate Missing Key Aspects Generation}
\label{sec:reasoning}

Utilizing samples from 30 cluster centers, we leverage the in-context learning of LLM to generate candidate answer. We employ different questioning methods based on the same samples to maintain consistency between candidate answers and samples. Moreover, we utilize a selector to choose answers from different questioning methods\cite{DBLP:journals/corr/abs-2209-01975, DBLP:journals/corr/abs-2301-00303}.

\begin{figure}[t]
    \centering

    \includegraphics[width=0.48\textwidth]{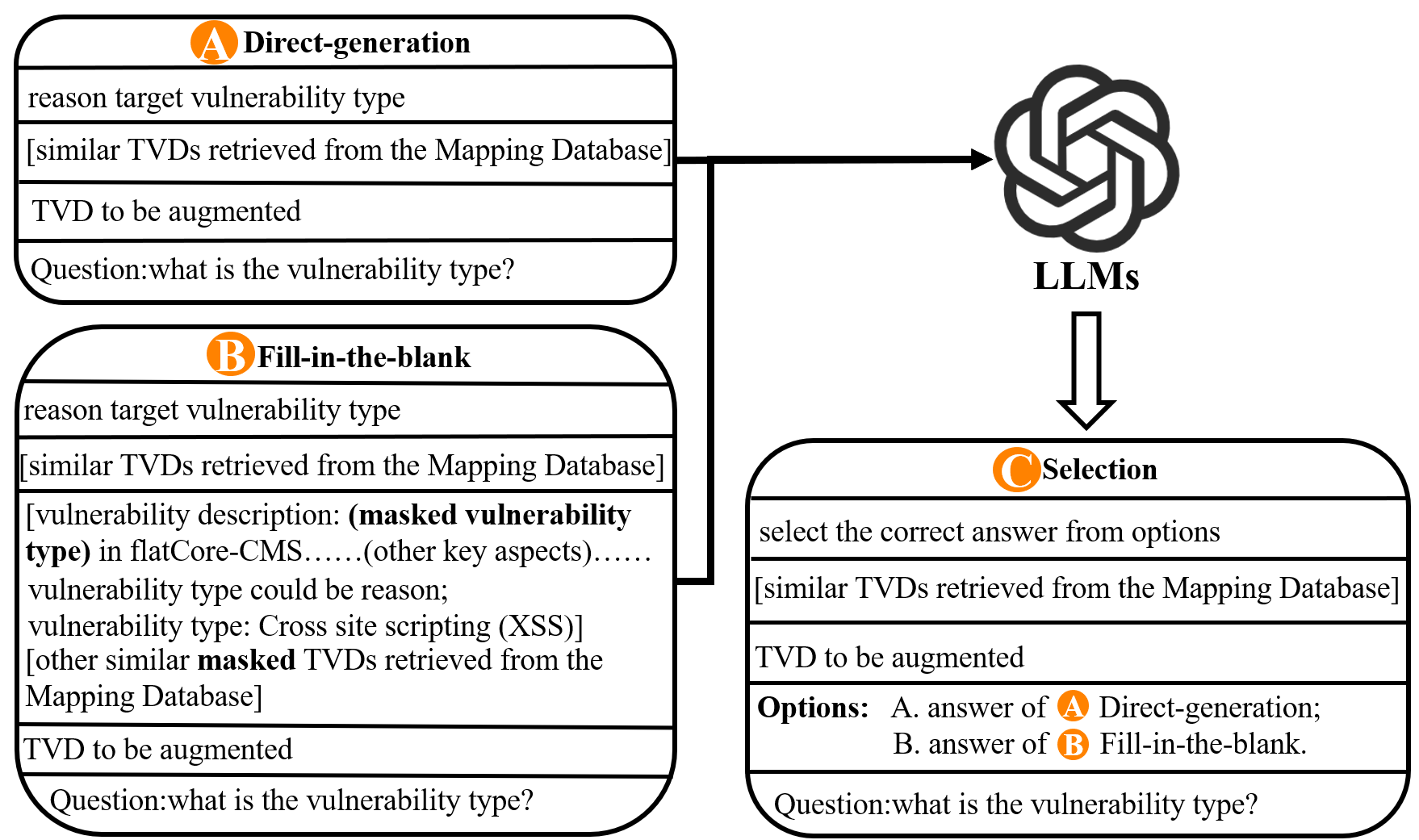}
    \caption{Prompt templates of missing key aspect generation.}
    \label{AI-Chain}

\vspace{-5mm}
    \end{figure}

In lines 12-14 of Algorithm \ref{alg:alg3}
, both the missing key aspects of the augmented TVD and the key aspects of the TVDs of the cluster centers are incorporated into the prompts ``GenerationStruct'', ``fillStruct'' and ``sellectStruct''. The final answer is determined through a selection prompt, including the two generated answers and the key aspect within its structure. Finally, Algorithm \ref{alg:alg3} is executed multiple times to generate multiple candidate answers, ensuring their diversity.

In Figure \ref{AI-Chain}, the ``Direct-generation'' encompasses a task description, the samples set, TVD, and the question. The sample set primarily comprises key aspects extracted from TVDs associated with 30 cluster centers. Each key aspect is structured in a key-value format, for instance, ``vulnerability\_type: Cross...''.
Meanwhile, the ``Fill-in-the-blank'' utilizes the mask mechanism inspired by transformers. The vulnerability type is masked and replaced with ``(mask)'', accompanied by a prompt with ``vulnerability type could be reason ...'' indicating the specific content of the masked vulnerability type.
The ``Selection'' introduces options for the selection structure, with the generator's answer as an option for LLM. This structured approach aids LLM in predicting relevant content, thereby enhancing learning efficiency.

\subsection{Answer Selection based on Software Feature Hallucination Detection}

LLMs are widely criticised for their propensity to produce hallucinations, indicating that these models may confidently generate incorrect answers instead of admitting their limitations by responding with ``I do not know''.
Furthermore, the phenomenon of hallucination occurs more frequently in tasks for which the pre-training dataset contains limited information, such as long-tail software.
Thus, to filter out incorrect answers generated by the LLM in the last module, we utilize software feature information from external source by employing the Natural Language Inference (NLI)~\cite{DBLP:conf/iclr/GongLZ18}.

NLI models demonstrate proficiency in analyzing correlations between two texts, namely, candidate missing key aspects and software features. However, software features often contain specialized terms specific to computing, which hold nuanced meanings distinct from everyday language. This inherent complexity presents challenges for NLI models to accurately interpret the true meaning of these terms based solely on their literal interpretation.
To address the challenge posed by specialized computing terms, researchers may consider replacing the text embedding model with a computing domain embedding model or adjusting the network structure using attention mechanisms to prioritize non-specialized terms\cite{DBLP:journals/corr/abs-2301-00303, DBLP:journals/corr/abs-2306-06427}. However, due to the high frequency and rich semantic content of specialized terms in the software vulnerability domain, disregarding them would result in substantial information loss.
Therefore, we integrate Wikipedia~\cite{wiki}, one of the most largest online encyclopedia, as background information for specialized nouns in software features to enrich the semantics of these terms. 
The Learning Deep Structured Semantic Models (DSSM)~\cite{DBLP:conf/cikm/HuangHGDAH13} are commonly employed in recommendation systems, embedding both the query history and document simultaneously as non-specialized and specialized information for NLI judgment.  
Inspired by this concept, we employ DSSM to embedding two different types of text: background information and specialized nouns.

\subsubsection{Construction of Software Feature Correlation Model}

We use BERT and LLaMA to form DSSM, the former for embedding software features and the latter for embedding background information. Fully connected layers connect these models to construct NLI model based on DSSM.

\begin{figure}[t]
    \centering
    \includegraphics[width=0.48\textwidth]{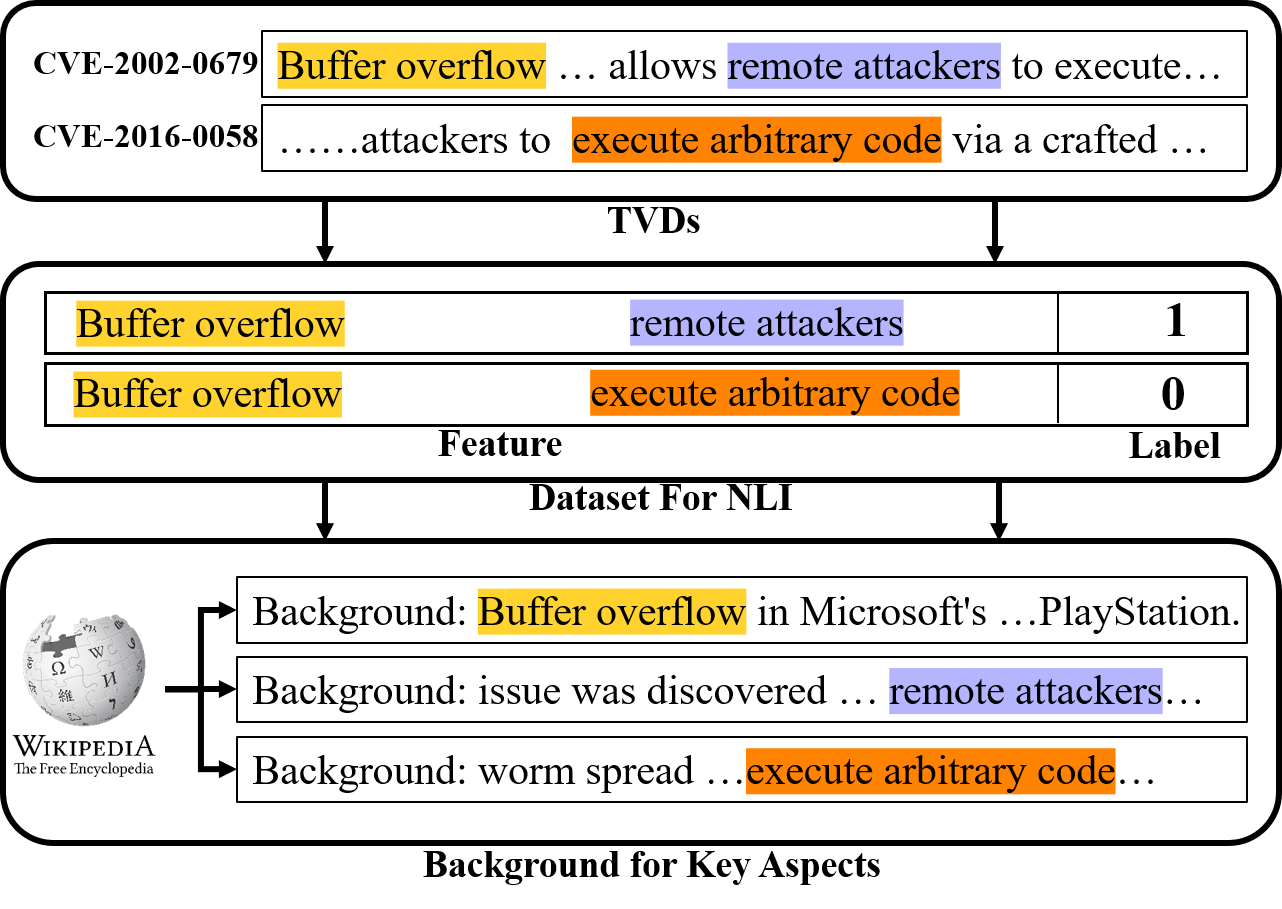}
    \caption{Training set construction for software feature correlation model.}
    \label{Training}

\end{figure}

All key aspects within a TVD are semantically related to each other. Thus, as shown in Figure \ref{Training}, pairwise extraction of key aspects within the same TVD forms the training set for the NLI model, with labels set to 1. Key aspects from different TVDs constitute the training set, with labels set to 0.
Given the significantly higher proportion of negative samples compared to positive samples, downsampling is employed to balance the negative sample ratio to twice that of the positive samples, ensuring the efficacy of the NLI model.
For each software feature (key aspect) in feature-text pairs, a keyword-based search is conducted in the WikiPedia~\cite{wiki} to retrieve the background information text.
This process results in a background information set of the same size as the feature-text pairs.
Ultimately, we constructe an NLI model capable of determining the relationship between two key aspects.

\subsubsection{Hallucination Key Aspect Filter}

\begin{figure}[t]
    \centering

    \includegraphics[width=0.48\textwidth]{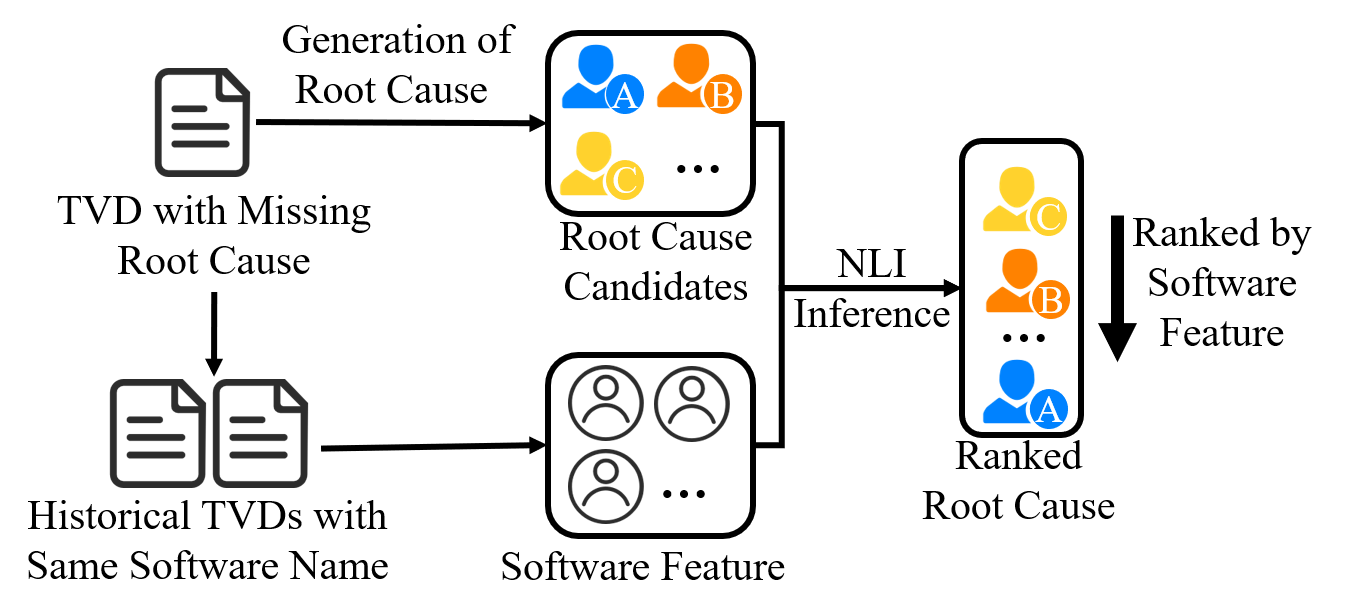}
    \caption{Hallucination key aspect filter.}
    \label{Selection}
    \vspace{-5mm}
    \end{figure} 

For each missing key aspect, as shown in Figure~\ref{Selection}, the \textit{Missing Key Aspect Generation} module will output multiple candidate answers.
Then, all candidate answers will be ranked according to their NLI probabilities, and the highest one will be taken as the final result of the generative prediction.
Specifically, first, we extract all software features (key aspects) corresponding to the TVD's software from the associated TVDs and calculate the correlation probabilities by the trained NLI model. Then, we aggregate the association probabilities between all features and the predicted results (Root cause1, ..., Root causeN). Finally, the prediction with the highest probability is the result.

%% file: 6_Evaluation.tex
\section{Evaluation}~\label{sec:setup}
We conduct a series of experiments to investigate the following research questions:

\begin{itemize} [leftmargin = *]

    \item RQ1: \textbf{[Missing Information Generation]} What is the performance of candidate missing key aspect generation?
    \item RQ2: \textbf{[Generated Answer Selection]} What is the performance of software feature hallucination detection-based answer selection?
    \item RQ3: \textbf{[Comparison]} What is the performance of comparison against baselines?
    \item RQ4: \textbf{[Downstream Tasks]} How can the augmented complete TVDs benefit downstream tasks? 
\end{itemize}

In this paper, we propose four research questions (RQs) to evaluate our augmentation framework for key aspects in long-tail software. RQ1 examines the performance of generating missing key aspects, validating our approach through statistical analysis and LLM application. RQ2 focuses on accurate key aspect prediction by detecting software feature hallucinations using an NLI model. RQ3 assesses the framework's generalization across various software types, while RQ4 evaluates the value of augmented TVDs in improving downstream tasks. Our analysis demonstrates the framework's significant advantages in key aspect reasoning.

\subsection{Dataset}
\label{sec:dataset}

\subsubsection{Data Collection}
The TVDs are collected from several reputable databases, including Mitre's CVE~\cite{CVE} and the National Vulnerability Database (NVD)~\cite{nvd}, which contain 233,456 and 231,454 records, respectively.
These repositories document software vulnerability information since 1999.
CVE and NVD are often applied for vulnerability analysis, such as~\cite{DBLP:conf/kbse/YitagesuXZ00H21, DBLP:journals/tosem/GuoCXLBS22, DBLP:journals/compsec/LiXZYC23, DBLP:conf/icse-encycris/OkutanM22, DBLP:conf/iri/SumotoKWTYFK22,DBLP:conf/uss/FengL0W0YZS19}. However, they suffer from the issues for vulnerability analysis based on LLMs: these datasets contain TVD repositories that are already present
in the training corpora of popular LLMs, leading to an
overestimation of the missing key aspect augmentation capabilities.
Therefore, to address the potential risk of data leakage and accurately reflect the capability of our framework, we implement the following measures:
According to \href{https://platform.openai.com/docs/models/gpt-3-5}{OpenAI’s official website}, the training data for GPT-3.5 is cut off as of April 2023. To ensure that our benchmark dataset contains code repositories that have not been seen during model training, we select TVD date after May 1, 2023. the CVE and NVD dataset comprises 63,000 and 240 records.
In RQ3, we find that the NVD dataset had limited volume, so we decide to extend the time frame and combine the NVD data from 2022 and 2024 as a third dataset, referred to as NVD*, with 381 records.

\subsubsection{Data Preprocessing}

In the NVD database, we adopt the data processing method described in PMA\cite{DBLP:journals/tosem/GuoCXLBS22}, utilizing the ``ANALYSE'' as the TVD and the ``MODIFIED'' as missing key aspects annotated by maintainers. This approach allows us to extract a total of 5,379 entries from NVD that have both ``MODIFIED'' and ``ANALYSE'' records.

In the CVE database, we extract five key aspects individually. To craft the testing set, we manually mask each key aspects from the TVD, leaving the remaining text as the testing data and the masked key aspect as the ground truth label. However, a challenge arises as the remaining sentence after removing a key aspect is often incomplete. For example, in CVE-2002-0679, if the ``Buffer overflow'' vulnerability type is removed, the resulting sentence ``in Common Desktop Environment (CDE) ToolTalk RPC database server….'' lacks a subject and is not a complete sentence, diverging significantly from the standard TVD format.
We notice that security management personnel frequently employ vague terms when describing CVEs with unknown key aspects. For instance, the TVD of CVE-2020-14274 reads, ``Information disclosure vulnerability in HCL…personal data via unknown vectors.'' Here, the phrase ``unknown vectors'' substitutes for the \textit{Attack Vector}. Inspired by this convention, we replace the deleted key aspect in the TVD with placeholders like ``unknown vulnerability type'', ``unknown attacker type'', etc., respectively. This ensures that the sentence structure remains intact and consistent across all TVDs, despite missing key aspects.

Finally, the sizes of the three datasets we used are as follows: the CVE dataset contains 63,000 records, the NVD dataset has 240 records, and the NVD* dataset includes 381 records.

\subsection{Evaluation Metrics}
\label{manual evaluation}

To evaluate labeled data, we adopt quantitative evaluation based on statistic analysis.
This involves iterating the entire dataset and assessing the results using following generation task evaluation metrics. Common evaluation metrics for generation tasks include BLEU \cite{DBLP:conf/acl/PapineniRWZ02}, METEOR \cite{DBLP:conf/acl/BanerjeeL05}, ROUGE \cite{rouge}, and BERTscore \cite{DBLP:conf/iclr/ZhangKWWA20}, and we only specifically choose BERTscore in RQ1, RQ2 and RQ3 for several reasons. Firstly, BERTscore excels at capturing semantic meaning and context similarity from a semantic understanding perspective. In contrast, BLEU, ROUGE, and METEOR may rely more on lexical and n-gram matching, potentially overlooking deeper semantic nuances. Secondly, due to the variable expression of TVDs, BERTscore is less sensitive to word order and phrasing variations compared to BLEU, ROUGE, and METEOR. The remaining three metrics will be available in the GitHub repository\cite{github} for further reference. For RQ4, we adopt F1-score as Metrics, following the evaluation method of the downstream task.

\begin{algorithm}[H]
\caption{Missing Key Aspect Generative Prediction.}\label{alg:alg2}
\begin{algorithmic}
\STATE 1:\quad Let $n \text{ be the missing key aspect name of TVD}$; 
\STATE 2:\quad Let $l \text{ be the missing key aspect of TVD}$ ;
\STATE 3:\quad $name \gets Software\_name\_extract(\text{TVD})$ ;
\STATE 4:\quad $prompt \gets \text{{``What is ''}} \ +l+ \ \text{{``of TVD? TVD:''}} \ +v$;
\STATE 5:\quad $prompt \gets prompt + \text{{``software name:''}} \ name$;
\STATE 6:\quad $prompt \gets prompt + \text{{``noting:''}} \ +l+ \ \text{{``is phrase.''}}$
\FOR{$i = 1$ \TO $3$}
    \STATE \hspace{-0.5cm} 7:\quad $t \gets RandomSample(\text{all\_TVD})$;
    \STATE \hspace{-0.5cm} 8:\quad  $prompt \gets prompt + \text{{``TVD:}} \ t.replace(t.n, \text{{unknown}} + n)\text{{''}}$;
    \STATE \hspace{-0.5cm} 9:\quad  $prompt \gets prompt + \text{{``n:}} \ t.l\text{{''}}$;
\ENDFOR
\STATE 10:\quad $mka \gets LLM(prompt)$;
\STATE 11:\quad $sim \gets BERTscore(mka, l)$;
\STATE \textbf{return}  $sim$
\end{algorithmic}
\end{algorithm}

In section \ref{sec:formative_reasoning} of formative study, our evaluation method, depicted in Algorithm \ref{alg:alg2}, first employs LLMs (GPT-3.5) to extract the TVD software name. Subsequently, a prompt is created, incorporating task requirements, TVDs, missing key asmpect names, and software names. Following this, three TVDs are randomly selected, key aspects are removed, and ``unknown'' is substituted in their place. The removed key aspects then act as labels for in-context learning. Finally, a prompt is inputted into LLMs (GPT-3.5), and the generated results are compared with the labels.
To address potential ambiguity in LLM-generated answers, constraints are imposed on the format of the answers within the prompt, as detailed in the sixth line of Algorithm \ref{alg:alg2}.

\subsection{Missing Key Aspects Generation (RQ1)}
\label{sec:RQ2}

We conduct a evaluation of the candidate missing key aspects generation. Initially, we perform a statistical analysis of the software-CVE mapping database to gauge the impact of different software name extraction methods on database entries. Subsequently, we assess the diversity of examples by different retrieval methods. Additionally, we evaluate the effectiveness of cluster to ensure representative sampling. Finally, we investigate the influence of different methods of using LLMs on our framework's overall performance.

\subsubsection{Statistical Analysis of the Software-CVE Mapping Database}
We conduct a statistical analysis of the Software-CVE Mapping database by counting the number of software entries after employing different software name extraction methods.
\begin{figure}[htbp]
\centerline{\includegraphics[width=1\linewidth]{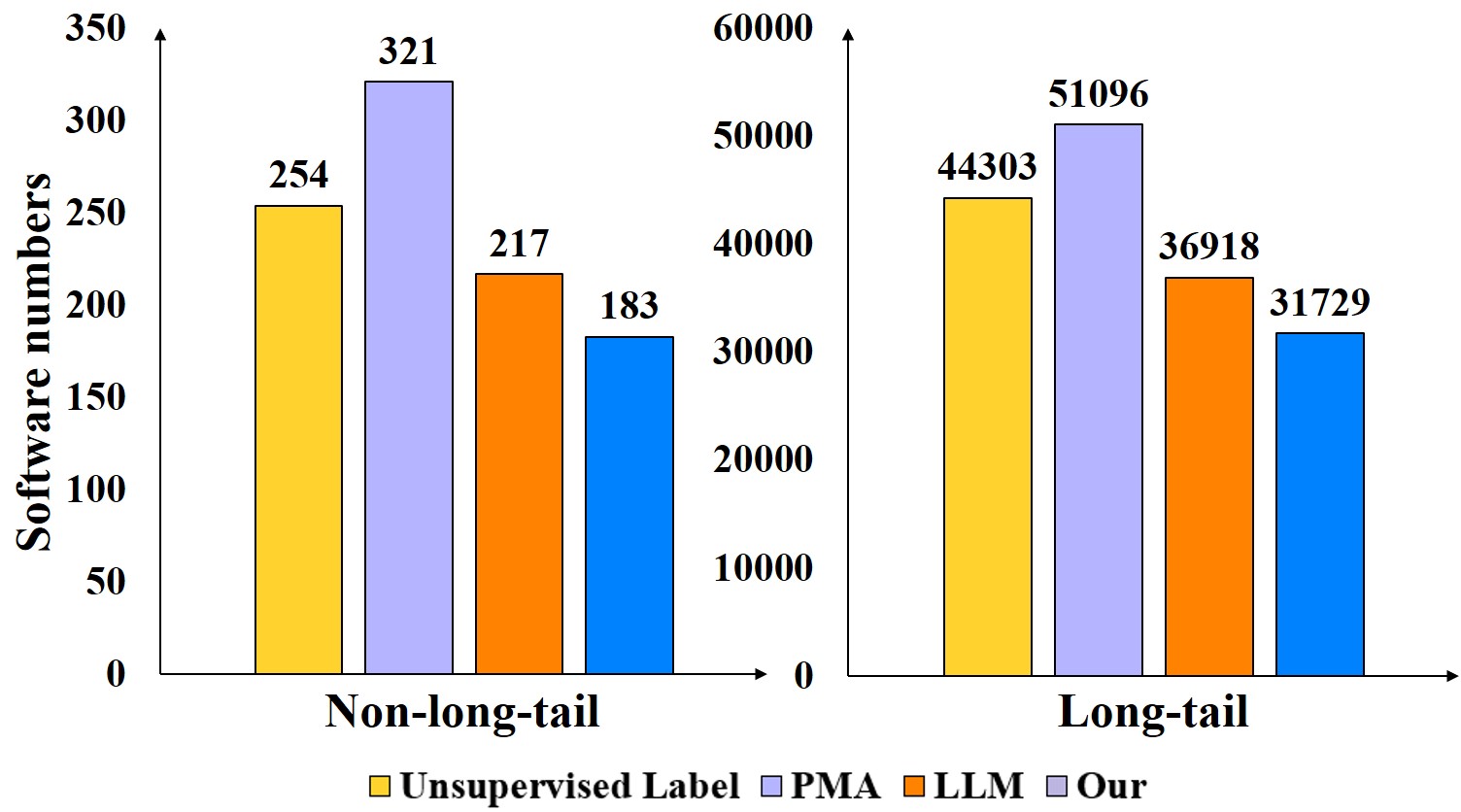}}
\caption{Statistical analysis for the number of software names. Unsupervised labeling involves BERT for key aspect extraction\cite{DBLP:conf/kbse/YitagesuXZ00H21}, while PMA\cite{DBLP:journals/tosem/GuoCXLBS22} regularizes existing key aspect extraction. The LLM prompt is ``What is the software name of the {TVD}.''}
\label{static numbers}
\end{figure}
In Figure \ref{static numbers}, our method reduces the number of software names, indicating that government-maintained repositories can reduce the inconsistency of software names. This reduction in inconsistency is particularly significant for long-tail software due to their higher prevalence of non-standard software names: the number of software names has decreased by 19,367 (51,096 - 31,729).

\subsubsection{Extraction of Key Aspects in TVD}
Our framework is fundamentally based on the extraction of key aspects, making the quality of key aspect extraction critical to the overall accuracy of the framework. In this paper, we leverage in-context learning, where key aspects are manually extracted from complete TVDs and used as examples for LLMs to learn and apply this process.

To assess the performance of our approach, we compare it with three relevant studies. Yitagesu et al.~\cite{DBLP:conf/kbse/YitagesuXZ00H21} develop an unsupervised method for labeling and extracting phrase-based concepts from vulnerability descriptions, improving the structured understanding of the text. Wei et al.~\cite{DBLP:journals/infsof/WeiBSLZT23} introduce an automated event extraction method for CVE descriptions, converting raw CVE text into structured events. Additionally, Yitagesu et al.~\cite{DBLP:conf/msr/YitagesuZ00X21} also propose a method for automatic part-of-speech tagging, which enhances linguistic processing by accurately tagging the parts of speech in vulnerability descriptions.
We follow the evaluation setup outlined in the work of Yitagesu et al.~\cite{DBLP:conf/kbse/YitagesuXZ00H21} , applying random sampling with a sample size of 4,000 entries to ensure robust analysis. The precision of key aspect extraction is evaluated using the F1-score, allowing for a comprehensive comparison between our framework and the related works.

\textcolor{red}{
\begin{table}[]
  \centering
  \caption{F1-score of Key Aspect Extraction. ``Our'' represents our proposed method, ``Unsuper.'' refers to the unsupervised learning method~\cite{DBLP:conf/kbse/YitagesuXZ00H21}, ``Event.'' denotes the event extraction method~\cite{DBLP:journals/infsof/WeiBSLZT23}, and ``Part.'' corresponds to the part-of-speech tagging approach~\cite{DBLP:conf/msr/YitagesuZ00X21} .}
  \label{ExtractionKeyaspect}
  \begin{tabular}{cccccc}
  \hline
    \tabincell{c}{Extraction\\Methods} & \tabincell{c}{Vulnerability\\Type}& \tabincell{c}{Attack\\Vector}& \tabincell{c}{Attacker\\Type}& \tabincell{c}{Impact}& \tabincell{c}{Root\\Cause}\\
    \hline
    \tabincell{c}{Our} & {\bfseries0.967} & {\bfseries0.959} & {\bfseries0.984} & {\bfseries0.948} &  {\bfseries0.953}\\
    \tabincell{c}{Unsuper.}& 0.898 & 0.876 & 0.899 & 0.868 & 0.878\\
    \tabincell{c}{Event.} & 0.884 & 0.868 & 0.979 & 0.861 & 0.897\\
    \tabincell{c}{Part.} & 0.875 & 0.857 & 0.893 & 0.836 & 0.841\\
    \hline    
  \end{tabular}
\end{table}
}

The results in the Table~\ref{ExtractionKeyaspect} demonstrate that our proposed method significantly outperforms existing approaches in extracting key aspects from vulnerability descriptions. With F1-scores consistently higher across all categories, particularly in \textit{Attacker Type} (0.984) and \textit{Vulnerability Type} (0.967), our method proves to be more accurate and reliable. In contrast, the unsupervised learning and event extraction methods show moderate performance, with noticeable gaps in critical areas like \textit{Vulnerability Type} and \textit{Root Cause}. The part-of-speech tagging approach lags behind in all categories, highlighting its limitations in capturing the complexity of technical aspects. Overall, this comparison underscores the effectiveness of our framework in key aspect extraction, crucial for improving vulnerability analysis.

\subsubsection{Diversity of Example Selection in In-context Learning}

In Table \ref{evidence_search}, we investigate the effect of retrieving various databases to enhance the diversity of TVDs.

\begin{table}[]
  \centering

  \caption{BERTscore of augmentation based on retrieving various databases. ``DB'' denotes retrieving only the software-CVE mapping database, ``DB\&CWE'' retrieving both the software-CVE mapping database and CWE, and ``No Retri.'' signifies not retrieving.}
  \label{evidence_search}

  \begin{tabular}{cccccc}
  \hline
    \tabincell{c}{Retrieving\\Methods} & \tabincell{c}{Vulnerability\\Type}& \tabincell{c}{Attack\\Vector}& \tabincell{c}{Attacker\\Type}& \tabincell{c}{Impact}& \tabincell{c}{Root\\Cause}\\
    \hline
    \tabincell{c}{DB\&CWE} & {\bfseries0.669} & {\bfseries0.593} & {\bfseries0.838} & {\bfseries0.625} &  {\bfseries0.581}\\

    \tabincell{c}{DB} & 0.601 & 0.522 & 0.771 & 0.537 &  0.529\\
    \tabincell{c}{No Retri.} & 0.283 & 0.137 & 0.580 & 0.175 & 0.159\\

    \hline    
   
  \end{tabular}

\end{table}
The ``DB\&CWE'' retrieving methods, which simultaneously utilizes software CVE mapping databases and CWE, consistently achieves the highest BERTscore values across all key aspects, indicating its effectiveness in enriching key aspects. This suggests that leveraging both software databases and CWE significantly enhances the diversity and accuracy of key aspects, leading to more robust augmentation.

\subsubsection{Representation of Example Selection}

Relevant and representative examples are crucial to achieve decent performance for LLM in-context learning.
We validate the effectiveness of our approach by comparing different example selection strategies, which include Kmeans, DBSCAN, OPTICS, sentence similarity-based retrieval, and random selection. We compare these methods to ensure that our example selection is representative: Kmeans is selected due to its efficiency and the clarity of its well-defined cluster centers, which ensures diversity of samples. Given that many key aspects in TVD are repetitive, such as in CVE-2002-0679 and CVE-2016-0058 shown in Figure~\ref{fig3}, leading to varying semantic densities, we utilize DBSCAN to leverage its ability to recognize different semantic densities. OPTICS, an improvement over DBSCAN, is used because it is less sensitive to clustering parameters, providing more stable clustering results. The BERT similarity-based method is a commonly used engineering approach that directly identifies similar sample examples. Random sampling serves as a baseline, highlighting the benefits of other methods and providing a straightforward comparison of the performance improvements achieved.

\begin{table}[]
  \centering

  \caption{BERTscore of Representative Sample Selection Methods. ``Sent. Sim.'' denotes retrieval based on sentence (BERT) similarity, and ``Random'' indicates random selection. }
  \label{Sample Selection}

  \begin{tabular}{cccccc}
  \hline
    \tabincell{c}{Selection\\Method} & \tabincell{c}{Vulnerability\\Type}& \tabincell{c}{Attack\\Vector}& \tabincell{c}{Attacker\\Type}& \tabincell{c}{Impact}& \tabincell{c}{Root\\Cause}\\
    \hline
    \tabincell{c}{Kmeans} & {\bfseries0.669} & {\bfseries0.593} & {\bfseries0.838} & {\bfseries0.625} &  {\bfseries0.581}\\

    \tabincell{c}{DBSCAN} & 0.601 & 0.522 & 0.771 & 0.537 &  0.529\\
    
    \tabincell{c}{OPTICS} & 0.590 & 0.491 & 0.773 & 0.523 & 0.542\\

    \tabincell{c}{Sent. Sim.}& 0.493 & 0.442 & 0.716 & 0.516& 0.407\\
    
    \tabincell{c}{Random} & 0.200 & 0.160 & 0.440 & 0.163 & 0.090\\

    \hline    
   
  \end{tabular}

\end{table}

Table \ref{Sample Selection} displays the BERTscore of various representative sampling strategies. Our kmeans method surpasses others, attaining the highest scores across all aspects, including \textit{Vulnerability Types}, \textit{Attack Vectors}, \textit{Attacker types}, \textit{Impact}, and \textit{Root causes}. Compared to clustering methods like DBSCAN and OPTICS, our approach exhibits superior performance, underscoring its efficacy in selecting representative examples. Conversely, strategies dependent on sentence similarity or random selection yield lower scores, underscoring the significance of systematic selection methods in vulnerability analysis.

\begin{table}
  \centering
  \caption{BERTscore of Different Number of Clusters}
  \label{optimal_number}

  \begin{tabular}{cccccccc}
    \hline
    \multicolumn{2}{c}{} & \multicolumn{6}{c}{\tabincell{c}{Number of Clusters}} \\
    \cline{3-8}
    \multicolumn{2}{c}{Key Aspect} & 5 & 10 & 20 & {\bfseries30} & 40 & 50\\
    \hline

    \multirow{5}{*}{} & Vuln. Type & 0.572 & 0.608 &0.651  &{\bfseries0.669} & 0.634 & 0.608\\
    & Attack Vector & 0.475 & 0.545 &0.584 & {\bfseries0.593} & 0.578 & 0.581\\
    & Attacker Type & 0.772& 0.803 & 0.836& {\bfseries0.838}& 0.831& 0.824\\
    & Impact & 0.559 & 0.568 & 0.604& 0.625 & {\bfseries0.629} & 0.592\\
    & Root Cause & 0.461 & 0.525 &0.565 & {\bfseries0.581}& 0.554&0.548\\
    \cline{1-2} \cline{3-8}
  \end{tabular}

\end{table}

Table \ref{optimal_number} indicates that the highest BERTscore is achieved for most aspects when the number of clusters is set to 30. Although the impact aspect shows the highest score at 40 clusters, the difference in accuracy between 40 and 30 clusters is negligible. Therefore, despite the slightly better performance observed at 40 clusters for impact, we choose to use 30 clusters as it offers comparable accuracy while maintaining simplicity in the model configuration.

\subsubsection{Application of LLMs in In-context Learning}

Our approach utilizes a three-question method for constructing the prompt. Therefore, we separately test the augmentation performance using a single prompt template and a combination of three questions.

\begin{table}[]
  \centering
  \caption{BERTscore of Prompt Strategy Combinations}
  \label{prompt_strategy}

  \begin{tabular}{cccc}
  \hline
    Key Aspect & {Direct} & {Fill-in}& {Gen+Fill+Selection}\\
    \hline
    {Vulnerability Type} &  0.551 & 0.545 &{\bfseries0.669}\\
    
    {Attack Vector}  &0.583& 0.514 &{\bfseries0.593}\\
    
    {Attacker Type} & 0.740 & 0.796 & {\bfseries0.838} \\
    
    {Impact} & 0.595 & 0.541 & {\bfseries0.625}\\
    
    {Root Cause}& 0.491 & 0.559 & {\bfseries0.581}\\
\hline
  \end{tabular}

\end{table}

In Table \ref{prompt_strategy}, the ``Gen+Fill+Selection'' combination consistently achieves the highest BERTscore values across all key aspects, indicating its effectiveness in augmenting missing key aspects. This suggests that employing a combination of generation, fill-in-the-blank, and selection strategies enhances the quality of candidate missing key aspects, leading to more accurate augmentation. Conversely, strategies relying solely on direct generation or fill-in-the-blank exhibit lower BERTscore values, underscoring the importance of utilizing comprehensive prompt strategies for missing key aspect augmentation.

Additionally, we compare the effectiveness of candidate answer generation by different LLMs, as depicted in Table \ref{different_LLM}.
\begin{table}[]
  \centering
  \caption{BERTscore of Different LLMs}
  \label{different_LLM}

  \begin{tabular}{ccccc}
  \hline
    Key Aspect & {LLaMA} & {T5}& {GPT-3.5}& {GPT4}\\
    \hline
    {Vulnerability Type} &  0.121 & 0.093 &{\bfseries0.669}&0.641\\
    
    {Attack Vector}  &0.076& 0.059 &0.593&{\bfseries0.607}\\
    
    {Attacker Type} & 0.573 & 0.549 & {\bfseries0.838} &0.819\\
    
    {Impact} & 0.095 & 0.108 & 0.625&{\bfseries0.643}\\
    
    {Root Cause}& 0.194 & 0.147 & {\bfseries0.581}&0.570\\
\hline
  \end{tabular}

\end{table}
GPT-3.5 consistently outperforms other models in augmenting \textit{Vulnerability Type}, \textit{Attacker Type}, and \textit{Root Cause}, achieving the highest BERTscore. On the other hand, GPT-4 achieves the highest BERTscore for \textit{Attack Vector} and \textit{Impact}, surpassing GPT-3.5 slightly. T5 and LLaMA also exhibit certain capabilities in augmenting \textit{Attacker Type}. This is attributed to the relatively limited categories that \textit{Attacker Type} can encompass, resulting in lower augmentation difficulty. However, T5 and LLaMA's poor performance across all key aspects is mainly due to their limited capability to understand lengthy prompts, leading to misinterpretations of questions and irrelevant responses. These results suggest that GPT-3.5 generally demonstrates superior performance across multiple key aspects compared to other LLMs.

\find{\textbf{Answer to RQ1:} The integration of software-CVE mapping database and CWE retrieval strategies of in-context learning significantly enhance the performance in missing key aspects generation.}

\subsection{Generated Answer Selection (RQ2)}
\label{RQ3}

Through experiments, we aim to determine 1) the optimal NLI model design for linking software features with key aspects and 2) evaluate the software feature hallucination detection-based answer selection method.

\subsubsection{Design of NLI Model}

\begin{table}[]
  \centering

  \caption{BERTscore of Different NLI Design}
  \label{NLI Design}

  \begin{tabular}{cccccc}
  \hline
    Structure & \tabincell{c}{Vuln.\\Type}& \tabincell{c}{Attack\\Vector}& \tabincell{c}{Attacker\\Type}& \tabincell{c}{Impact}& \tabincell{c}{Root\\Cause}\\
    \hline
    \tabincell{c}{LSTM+LSTM} & 0.597 & 0.524 & 0.783 & 0.574 &  0.522\\
    \tabincell{c}{BERT+BERT} & 0.621 & 0.553 & 0.801 & 0.599 &  0.547\\

    \tabincell{c}{RoBER+RoBER} & 0.639 & 0.570 & 0.817 & 0.594 &  0.568\\
    \tabincell{c}{LLaMA+LLaMA} & 0.644 & 0.581 & 0.813 & 0.617 & 0.560\\
    \tabincell{c}{LSTM+LLaMA} & 0.651 & 0.574 & 0.827 & 0.608 & 0.577\\
    \tabincell{c}{BERT+LLaMA} & {\bfseries0.669} & {\bfseries0.593} & {\bfseries0.838} & {\bfseries0.625} &  {\bfseries0.581}\\

    \hline    
   
  \end{tabular}

\end{table}

Table \ref{NLI Design} shows that the BERT+LLaMA combination achieves the highest BERTscore across all key aspects, with notable improvements over other model designs. For instance, in vulnerability type prediction, BERT+LLaMA attain a BERTscore of 0.669, outperforming other configurations by a significant margin. Similarly, in \textit{Attacker Type} prediction, BERT+LLaMA achieves a BERTscore of 0.838, demonstrating its superiority over alternative model architectures. These results underscore the effectiveness of leveraging advanced language representation models, like BERT and LLaMA, in enhancing the accuracy of key aspect prediction tasks related to software features.

\subsubsection{Selection of Top N}
\begin{table}[]
  \centering
  \caption{Average BERTscore of the Top N candidate missing key aspects}
  \label{answer_ver}

  \begin{tabular}{ccccc}
  \hline
     Key Aspect& \tabincell{c}{Top 1}& \tabincell{c}{Top 3}& \tabincell{c}{Top 5}\\
    \hline
    \tabincell{c}{Vulnerability Type}&{\bfseries0.669}& 0.631 & 0.617\\     
    
    \tabincell{c}{Attack Vector}&  {\bfseries0.593} &0.551&0.525\\
    
   \tabincell{c}{Attacker Type} &   {\bfseries0.838} &0.833&0.826\\
    
   Impact & {\bfseries0.625} &0.589 &   0.532 \\
    
  \tabincell{c}{Root Cause}&{\bfseries0.581}  &0.563  &  0.528 \\

    \hline
  \end{tabular}

\end{table}

Table \ref{answer_ver} presents the average BERTscore of the top N candidate missing key aspects across different key aspects. We observe a decreasing trend in BERTscore as the number of candidate answers increases from the top 1 to the top 5. This trend suggests that the accuracy of augmenting missing key aspects tends to decrease as more candidate answers are considered and our team effectively detect the hallucinatory answer. However, the specific impact varies across key aspects. For instance, the BERTscore for \textit{Attacker Type} remains consistently high across all levels (top 1, 3, and 5), indicating the degree of hallucination in the \textit{Attacker Type} being relatively low. In contrast, the BERTscore for \textit{Impact} exhibits a more significant drop from the top 1 to the top 5, suggesting greater difficulty in accurately augment this key aspect. Overall, these findings highlight the importance of considering the number of candidate missing key aspects in the answer selection process to optimize prediction accuracy.

\find{\textbf{Answer to RQ2:} Our selection method of candidate missing key aspects can effectively detect LLM hallucinations and select the correct key aspect answer.}

\subsection{Comparison Against Baselines (RQ3)}
\label{sec:RQ4}

In this section, we evaluate the performance of our proposed method across multiple datasets. Additionally, we also aim to evaluate our method's impact on non-long-tail software accuracy while enhancing accuracy for long-tail software. 

Our baseline adopt the method setting from Algorithm~\ref{alg:alg2}, with CVE dataset, using few-shot learning for comparison in augmentation. To enhance the objectivity of the engineering comparison, we include GPT-3.5, GPT-4 and LLaMA as LLMs in Algorithm~\ref{alg:alg2}. In real-world engineering, software engineers often use LLMs under three settings: 1) prioritizing the highest accuracy without regard to cost, and 2) prioritizing low cost. GPT-4 represents the former, with its high accuracy, while LLaMA represents the latter, as it can be deployed locally at no cost, and 3) balancing accuracy and cost. GPT-3.5 offers a compromise between performance and expense, though it may not excel in either area. 
Algorithm~\ref{alg:alg2} employs few-shot learning to effectively leverage the LLM's generative capabilities.
Additionally, we also compare the performance across different datasets. We evaluate the baseline performance on three datasets: CVE, NVD, and NVD*.

\begin{figure*}[htbp]
    \centering
    \includegraphics[width=1\textwidth]{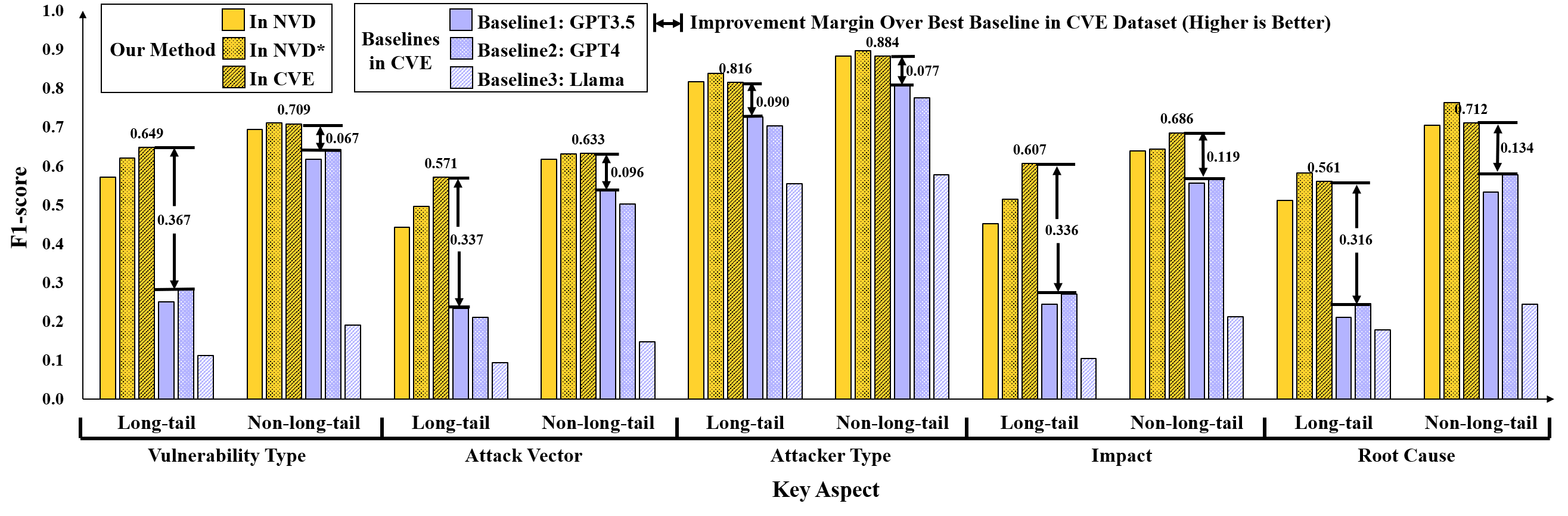}
    \caption{F1-Score comparison of different augmentation methods across Long-tail and Non-long-tail software in different dataset. \raisebox{-0.2ex}{\includegraphics[height=0.3cm]{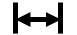}} with improvement margin values indicate that our method achieves a \textbf{significantly greater improvement} for long-tail software compared to non-long-tail software.}
    \label{RQ3}

    \end{figure*}

In Figure~\ref{RQ3}, our method demonstrates robust performance across all datasets. For long-tail software, our approach shows a clear advantage over baseline methods. While models like GPT-4 and LLaMA struggle significantly (e.g., scoring as low as 0.112 for \textit{Vulnerability Type} on LLaMA), our method achieves much higher and more consistent scores. On average, our method improves performance for long-tail software by 0.236 compared to the best baseline, indicating its ability to handle the inherent complexity and sparsity of long-tail data, which poses challenges for other models.
In contrast, for non-long-tail software, the performance gap between our method and the baselines narrows, as these datasets provide richer and more uniform features. Our method achieves an average improvement of 0.078 for non-long-tail software. Nevertheless, our approach maintains competitive results, as demonstrated by a score of 0.712 for \textit{Vulnerability Type} on NVD*, which aligns closely with the baseline CVE score of 0.709.
The distinct advantage of our method in long-tail scenarios underscores its capability to generalize effectively in more challenging contexts, where baseline LLMs often falter due to the lack of representative patterns. The observed difference in average improvement—0.236 for long-tail compared to 0.078 for non-long-tail—highlights the significant contribution of our framework to long-tail software. This makes our method a superior choice for applications where data diversity and sparsity are prevalent.

However,  we consistently observe lower accuracy when augmenting the key aspects \textit{Attack Vector}, \textit{Impact}, and \textit{Root Cause}, especially in comparison to \textit{Attacker Type} and \textit{Vulnerability Type}. This discrepancy primarily stems from two factors:
\begin{itemize}
    \item \textbf{Increased Diversity of Possible Values}: For \textit{Attack Vector}, potential values range from simple network attacks to complex user interactions. For instance, CVE-2017-0144 involves the EternalBlue exploit via network-based delivery, while CVE-2021-34527 requires user interaction. This broad variability makes prediction more difficult, unlike more straightforward aspects like \textit{Attacker Type} (e.g., remote vs. local). Similarly, \textit{Impact} and \textit{Root Cause} also exhibit a wide range of outcomes, such as CVE-2019-0708's system compromise and CVE-2020-0601's integrity violation, increasing the complexity for accurate predictions.
    \item \textbf{Longer Content}: The length of content for certain key aspects also impacts performance. For example, \textit{Root Cause} often involves detailed explanations, such as CVE-2018-11776’s Apache Struts vulnerability caused by input validation issues. In contrast, \textit{Attacker Type} or \textit{Vulnerability Type} are typically more concise. The model's accuracy drops as it struggles with longer input text, particularly for content-heavy aspects like \textit{Root Cause}.
\end{itemize}
The limitations observed in our framework are largely due to the inherent nature of the data itself, rather than the methodology. The high variability in key aspects like \textit{Attack Vector} and the length of detailed explanations for \textit{Root Cause} create challenges that affect any model's performance. These issues stem from the complexity and diversity of real-world vulnerability descriptions, making it difficult to maintain consistent accuracy across all key aspects.

Reviewing the examples curated in Figure~\ref{fig3}, our framework can successfully augment the missing \textit{Attack Vector} for CVE-2002-0679 and CVE-2016-0058 as ``Manipulating the \_TT\_CREATE\_FILE procedure argument'' and ``Exploiting the PDF Library via a malicious PDF document'', respectively. 

\find{\textbf{Answer to RQ3}: The proposed framework demonstrates robust performance across multiple datasets, showcasing satisfying generalization capabilities.}

\subsection{Augmentation on Different Tasks (RQ4)}
\label{sec:RQ5}

In the field of software maintenance and software vulnerability information management, typical applications are software vulnerability level (CVSS) prediction\cite{DBLP:conf/wcre/LiRXXS23}, software vulnerability CWE category prediction\cite{DBLP:journals/compsec/WangGRZ23}, and software identification of vulnerability libraries\cite{DBLP:conf/iwpc/HaryonoK0SSA022}. CVSS prediction involves classifying the severity of vulnerability information, while CWE focuses on clustering and merging vulnerability information. Dependency library identification aims to recognize libraries that are not described in TVD but actually impacting the system (library). These tasks effectively manage and categorize vulnerability, enabling security personnel to efficiently review and archive vulnerabilities.
They rely on the quality of TVD. Our work focuses on enhancing the quality of TVD, aligning with the requirements of these tasks.

For the CVSS prediction task, following the approach of Li et al.~\cite{DBLP:conf/wcre/LiRXXS23}, we utilize CVE dataset and extract key aspects, which are tokenized using a pre-trained BERT model to generate semantic representations. The task is treated as a regression problem where the model predicts normalized CVSS Base Scores. The dataset is split into 80\% training, 10\% validation, and 10\% testing, with performance evaluated using F1-score.
For CWE prediction, similar to Wang et al.~\cite{DBLP:journals/compsec/WangGRZ23}, the CVE dataset is used to classify vulnerabilities into CWE categories. One-hot encoding is applied to the multi-label CWE tags. A BERT-based model is fine-tuned for multi-label classification. The dataset is split into 70\% training, 15\% validation, and 15\% testing, with F1-score metrics.
For libraries identification, as outlined in Haryono et al.~\cite{DBLP:conf/iwpc/HaryonoK0SSA022}. We use the CVE dataset to classify vulnerabilities based on the affected software libraries from 381 categories. Key aspect are extracted, and tokenization is performed with a transformer model (BERT). Oversampling and undersampling are applied to address label imbalance, and the dataset is split into 60\% training, 20\% validation, and 20\% testing. F1-score is used for evaluation.
Overall, across all tasks, we utilize advanced deep learning techniques and follow state-of-the-art methodologies for each task to ensure the robustness of our results. Extensive hyperparameter tuning is also conducted to optimize model performance.

\begin{table}[]
  \centering
  \caption{F1-score of Generalizing Across Different Tasks}
  \label{muti task}

  \begin{tabular}{cccc}
  \hline
    Task Name & {Original} & {Augmented}\\
    \hline
    {CVSS Prediction} &  0.812 & {\bfseries0.857}\\
    
    {CWE Prediction}  &0.911& {\bfseries0.930}\\
    
    {Libraries Identification} & 0.786 & {\bfseries0.799}\\
\hline
  \end{tabular}

\end{table}

The Table \ref{muti task} presents the F1-score of generalizing across different tasks, comparing the performance of the original method with the augmented TVDs. Across all tasks, the method with the augmented TVDs consistently outperforms the original, indicating the effectiveness of our augmentation approach in improving performance across diverse tasks.

We observe that for the libraries identification task, the improvement in accuracy is less pronounced, increasing from 0.786 to 0.799. This modest improvement highlights the primary challenge of this task, which lies in the large number of classification labels. In our dataset, there are 381 distinct libraries, each serving as a classification label. This vast number of categories introduces significant complexity for the model in several ways. First, the label overhead and complexity increase as the model must differentiate between a large set of libraries based on often subtle differences in vulnerability descriptions. Libraries within similar ecosystems, such as various web frameworks, may have vulnerabilities described in similar ways, making classification more challenging. With a higher number of labels, the risk of misclassification rises, particularly when many libraries are underrepresented. Second, there is a significant class imbalance in the dataset, where some popular libraries have numerous associated vulnerabilities while others are sparsely represented. This imbalance leads to overfitting on frequent labels and reduces the model’s effectiveness on underrepresented categories. Given these challenges, the observed improvement in accuracy, although seemingly modest, is significant. 
In addition, we find that for the libraries identification task, small improvements are generally acceptable due to the large number of labels. Haryono et al.~\cite{DBLP:conf/iwpc/HaryonoK0SSA022} conduct a study on different methods for the libraries identification task and observe that even with various approaches, the improvements remain quite limited. Specifically, the paper surveys models like DiSMEC, XML-CNN, FastXML, ExtremeText, Parabel, Bonsai, and LightXML, and report an average accuracy improvement of only 1.4\% for each more advanced method. Therefore, our 1.3\% accuracy improvement is reasonable and acceptable.

\find{\textbf{Answer to RQ4:} Our method augment the key aspects of TVDs, improving the performance of downstream tasks that utilize TVD data.}

%% file: 7_Discussion.tex
\section{Discussion}

\subsection{Analysis of Augmented Impact}
\label{Analysis of Augmented Impact}
We explore the benefits of accurate augmentation and the negative impact of incorrect augmentation, and explore their boundaries. The purpose of vulnerability information management is to assist software maintenance personnel in current vulnerability repair. In this scenario, assuming a software has a vulnerability and the TVD lacks a key aspect. If it is necessary to know the missing key aspect, software maintenance personnel will manually supplement the key aspect by consulting the vulnerability information database. 
Assuming that software maintenance personnel find n pieces of relevant TVD as references through turn to external resources or internal search.
software maintenance personnel will review $n$ times to obtain vulnerability information, that is, the expected number of times software maintenance personnel review is:
\begin{equation}
\small
E (\text{not\_augment})=n
\end{equation}
Assuming that our method is used for key aspect augmentation, and our method accuracy is $x$. The probability of software maintenance personnel obtaining the missing key aspect only once is $x$, and a probability obtaining missing key aspect through $(1+n)$ times is $(1-x)$. $(1+n)$ shows that the error key aspect augmented is first attempted, and then $n$ records are attempted. For augmented missing key aspect, the expected number of views for software maintenance personnel is:
\begin{equation}
\small
E (\text{augment})=x+(1-x) (1+n). 
\end{equation}
When $E (\text{augment})<E (\text{not\_augment})$, we believe that our approach will be helpful in practice. Therefore, when 
\begin{equation}
\small
x + (1-x)(1+n) < n \Rightarrow x > \frac{1}{n}
\end{equation}
our algorithm is effective. When $n=[0,1]$, the augmentation algorithm is not applicable to practice. When $n>1$, as long as the accuracy of the augmentation algorithm is greater than 0.5, it will save more query times than manual work.

In addition, It is possible to mandate that users provide all key aspects when submitting TVD. But sometimes users themselves may not be aware of certain key aspects, and enforcing will result in delayed or abandoned submission due to users' unwillingness to invest additional time.

\subsection{New Paradigm for Information Augmentation tasks in Software Engineering}
With the development of LLMs, information augmentation tasks in software engineering are increasingly inclined to utilize LLMs as knowledge repositories for generating new information. However, the richness of information in LLMs can result in hallucinations in the generated results, leading to a lack of trust in the augmentation outcomes.

The current paradigm for addressing this issue involves using knowledge repositories to detect hallucinations at the level of text similarity. However, a major challenge of this approach is that the knowledge repository may offer multiple pieces of evidence supporting different augmentation outcomes.

At the heart of software engineering lies the ``\textbf{software}'' itself. Our new paradigm proposes focusing on each software individually for information augmentation and hallucination detection. Our framework, based on detecting hallucinations for each software, aims to systematically uncover the unique characteristics of each software, thereby enhancing the robustness of information augmentation.

\subsection{The Impact of Different Long-tail and Non-long-tail Thresholds on Software Feature Inference Framework}
We solely employ CVE and BERTscore as datasets and evaluation metrics to observe the impact of various thresholds on our augmentation results.

\begin{table}[]
  \centering
  \caption{BERTscore of Different Long-tail Threshold. ``NL'' represents non-long-tail software, and ``L'' denotes long-tail software.}
  \label{Threshold}
  \begin{tabular}{c|cccccc}
  \hline
   Threshold&\tabincell{c}{Soft. \\Type}&\tabincell{c}{Vuln.\\Type}& \tabincell{c}{Attack\\Vector}& \tabincell{c}{Attacker\\Type}& \tabincell{c}{Impact}& \tabincell{c}{Root\\Cause}\\
    \hline
    \multirow{2}{*}{25} & \textbf{L} & 0.632&0.518 &0.761  &0.587  &  0.509 \\
     & \textbf{NL} & 0.611&0.495 &0.851  &0.526  &  0.589 \\
    \hline
    \multirow{2}{*}{50}  & \textbf{L}& {\bfseries0.649}& 0.571& {\bfseries0.816} & 0.607 & {\bfseries0.571}  \\
     & \textbf{NL} & 0.709&0.633 &0.884  &0.686  &  0.712 \\
    \hline
    \multirow{2}{*}{75}  & \textbf{L} & 0.648& 0.575 & 0.814 & {\bfseries0.608} & {\bfseries0.571} \\   
     & \textbf{NL} & 0.711&0.639 &0.887  &0.698  &  0.720 \\
    \hline
        \multirow{2}{*}{100}  & \textbf{L} & 0.646& 0.581 & 0.810 & 0.592 & 0.569 \\  
     & \textbf{NL} & {\bfseries0.715}&0.641 &0.887  &0.701  &  0.712 \\
    \hline
        \multirow{2}{*}{125}  & \textbf{L} & 0.649& {\bfseries0.588} & 0.811 & 0.605 & 0.570 \\
     & \textbf{NL}& 0.706&0.624 &0.871 &0.705  &  0.708 \\
    \hline
  \end{tabular}
\end{table}

The Table \ref{Threshold} illustrates BERTscore results across different long-tail thresholds (ranging from 25 to 125), highlighting a notable deviation at a threshold of 25, where BERTscore values exhibit a slight decrease. However, from thresholds 50 to 125, the differences in BERTscore values between long-tail (L) and non-long-tail (NL) software categories are negligible. Specifically, for long-tail software, BERTscore values remain relatively stable, ranging from 0.64 to 0.65 across the thresholds. Conversely, in the non-long-tail category, BERTscore shows a consistent increase from 0.61 at threshold 25 to 0.71 at threshold 100. This indicates that while there's a distinct performance pattern at a threshold of 25, there's minimal variation in BERTscore values across thresholds 50 to 125, suggesting consistent model performance within this range for both long-tail and non-long-tail software.

\subsection{Threats to Validity}

\subsubsection{Internal Validity} 
As far as we know, we adopt the best method for extracting software names. But it is obvious that the method is less accurate than manual extraction, which is not realistic due to the large amount of data. Therefore, a bottleneck that restricts our method is TVD software name extraction accuracy.
    
\subsubsection{External Validity}
We acknowledge that several factors may limit the external validity of our study. First, our data is collected from specific sources (CVE and NVD). The two datasets may not fully represent the entire population of software vulnerabilities. Second, the study only evaluation the effectiveness of our method on a specific set of metrics (BERTscore) and may not apply to other performance metrics. Despite these limitations, our findings contribute to missing key aspect augmenting and can provide valuable insights for future research.

\subsection{Limitations}
Our approach, while effective in many respects, faces several challenges that stem primarily from the data quality and the inherent limitations of transformer-based models like GPT.
\begin{itemize}
    \item \textbf{Vague Key Aspect Descriptions}: Datasets like CVE and NVD often provide vague descriptions of key aspects such as \textit{Attack Vector} or \textit{Root Cause}. For instance, CVE-2021-34527 lacks specific details on how the attack is executed compared to more detailed entries like CVE-2017-0144. This vagueness impairs the model’s ability to generate precise predictions.
    \item \textbf{Model Challenges}: GPT-based models handle simpler aspects like \textit{Attacker Type} well but struggle with more complex, variable fields such as \textit{Impact} or \textit{Attack Vector}. For instance, predicting straightforward network vulnerabilities is easier than addressing complex scenarios involving multiple steps or user interactions, which can lead to decreased performance for intricate cases.
    \item \textbf{Lack of Qualitative Evaluation}: While our model performs well in quantitative metrics, it lacks human validation. For complex vulnerabilities like CVE-2018-11776, human expertise is necessary to ensure the generated explanations accurately reflect the underlying issues.
\end{itemize}
Future work will focus on developing a comprehensive TVD evaluation framework. This framework aims to systematically assess the quality and effectiveness of TVDs in software remediation. It will involve creating quantitative metrics to evaluate how well TVDs capture essential details and their impact on practical remediation efforts. By incorporating diverse evaluation criteria, including completeness, consistency, and practical relevance, this framework will provide a more nuanced understanding of how TVDs contribute to effective software repair. This approach will help identify areas where TVDs can be improved and ensure that the generated TVDs are not only accurate but also useful in real-world scenarios.

%% file: 9_Related_work.tex
\section{related work}

\textbf{Software Vulnerability Information Mining}: Vulnerability reports have been widely used for the research of Vulnerabilities discovering and analyzing or Vulnerabilities protection\cite{DBLP:conf/eurosp/GeTPJ16, DBLP:conf/uss/BiswasFCRVNFP17, DBLP:conf/uss/FengL0W0YZS19, DBLP:conf/iceccs/GongXLFH19, DBLP:conf/icsm/HanLXLF17, DBLP:conf/uss/LuPW19, DBLP:journals/tissec/PomonisPKPK18, DBLP:conf/ndss/WuHML20}. As a result, generating and augmenting high-quality vulnerability reports has been the subject of extensive research in recent years. Kensuke et al. \cite{DBLP:conf/iri/SumotoKWTYFK22} and Hattan et al. \cite{DBLP:journals/corr/abs-2210-01260} have employed deep learning methods to enrich vulnerability reports for non-long-tail software by learning patterns and features from large-scale training data. However, these methods may not be effective for long-tail software due to their unique characteristics and limited training data.

\textbf{TVD Key Aspects Extraction and Augmentation}: Many studies have proposed natural language processing (NLP) techniques to extract key information from vulnerability reports and classify them by severity or priority \cite{DBLP:conf/kbse/YitagesuXZ00H21, DBLP:conf/icse-encycris/OkutanM22,DBLP:conf/raid/0003YLOKR20, DBLP:conf/acl/ChoiCL22a, DBLP:conf/msr/YitagesuZ00X21, DBLP:journals/infsof/WeiBSLZT23}. Also, NLP methods have been widely applied to software text (e.g. bug report, code comments, etc.) and code\cite{DBLP:conf/icse/HassanW17, DBLP:conf/semweb/QiuZQZLRRQTY18, DBLP:conf/sigir/RuWQZL0020, DBLP:conf/msr/ShokripourAKZ13, DBLP:journals/access/ZhangXWLLWZH20, DBLP:conf/sigsoft/Zhao0BSWMW20}.
For key aspect augmentation, Guo et al.~\cite{DBLP:journals/tosem/GuoCXLBS22}  identify a significant issue with missing key aspects and propose a classification-based approach for key aspect prediction, modeling the aspects as labels and predicting those labels. However, no existing research directly tackles key aspect prediction through generative methods.

\textbf{LLM Applications on Software Engineering}: LLMs has been adopted in various fields, including machine learning and cognitive science, to discover commonalities among existing data \cite{DBLP:journals/corr/abs-2201-11903}. It has been applied to text prediction tasks in specific domain, such as scientific abstracts and medical reports \cite{DBLP:journals/corr/abs-2302-06474, DBLP:journals/corr/abs-2207-08143}. Specifically, LLM is proven to have promising performance in Software Engineering\cite{DBLP:journals/corr/abs-2301-08653, DBLP:journals/corr/abs-2304-07232, DBLP:journals/corr/abs-2303-07839}. However, prior research has yet to explore the potential of LLM generation for vulnerability report generation for long-tail software. 

Therefore, we propose a approach utilizing LLM generation to augment missing key aspects in long-tail software. The method based on AI chain and product-centric database prediction could reduce existing methods' limitations in domain-specific tasks with little and imbalanced data. Moreover, our approach explores a new paradigm of vulnerability report generation, which may enhance the vulnerability resolution process, ensuring safe use in critical applications.

%% file: 8_Conclusion.tex
\section{conclusion}
The TVDs for vulnerability are often incomplete on the CVE, and existing research on missing information augmentation perform poorly on long-tail software.
The issue stems from the lack of sufficient information in TVDs for long-tail software, resulting in challenges when utilizing TVDs for various tasks, such as CVSS and CWE prediction.
To address this limitation, we propose a software feature inference framework to augment missing TVD information by leveraging software features derived from software history data.
The evaluation shows that our solution can effectively generate missing information and select the correct answer by dispelling LLM hallucinations. 
Our framework also demonstrates satisfying generalizability and decent performance on downstream tasks.

%% file: 11_Acknowledge.tex
\section*{Acknowledgment}
This work was supported by the Open Project of CHINA ACADEMY OF RAILWAY SCIENCES CORPORATION LIMITED (RITS2023KF05).